\documentclass[namedreferences]{solarphysics}
\usepackage[optionalrh]{spr-sola-addons} 
\usepackage{graphicx}        
\usepackage[abs]{overpic}
\usepackage{color}           
\usepackage{url}             

\usepackage{lscape}

\usepackage{here}

\usepackage{hyperref}
\ifx \doiurl    \undefined \def \doiurl#1{\href{http://dx.doi.org/#1}{\textsf{DOI}}}\fi
\ifx \adsurl    \undefined \def \adsurl#1{\href{http://adsabs.harvard.edu/abs/#1}{\textsf{ADS}}}\fi
\ifx \arxivurl  \undefined \def \arxivurl#1{\href{http://arxiv.org/abs/#1}{\textsf{arXiv}}}\fi

\newcommand{\etal}{{\it et al.}}

\newcommand{\aap}{    {\it Astron. Astrophys.}}

\newcommand{\apj}{    {\it Astrophys. J.}}
\newcommand{\apjl}{   {\it Astrophys. J. Lett.}}

\newcommand{\pasj}{   {\it Pub. Astron. Soc. Japan.}}
\newcommand{\pasjl}{   {\it Pub. Astron. Soc. Japan. Lett.}}

\newcommand{\solphys}{{\it Solar Phys.}}

\begin{document}

\begin{article}

\begin{opening}

\title{Evolution and Flare Activity of $\delta$-Sunspots in Cycle 23}
\author{K.~\surname{Takizawa}$^{1}$\sep
        R.~\surname{Kitai}$^{1, 2}$
       }
\runningauthor{Takizawa and Kitai}
\runningtitle{Evolution and Flare Activity of $\delta$-Spots in Cycle 23}

   \institute{$^{1}$ Kwasan and Hida Observatories, Kyoto University, Yamashina-ku, Kyoto 607-8417, Japan\\
                $^{2}$ Bukkyo University, Kita-ku, Kyoto 603-8301, Japan\\
                     email: \url{takizawa@kwasan.kyoto-u.ac.jp} \\ 
                     email: \url{kitai@kwasan.kyoto-u.ac.jp}
             }

\begin{abstract}
The emergence and magnetic evolution of solar active regions (ARs) of $\beta\gamma\delta$-type, which are known to be highly flare-productive, were studied with the SOHO/MDI data in Cycle 23.
We selected 31 ARs that can be observed from their birth phase, as unbiased samples for our study.
From the analysis of the magnetic topology (twist and writhe), we obtained the following results.
\romannumeral1 ) Emerging $\beta\gamma\delta$ ARs can be classified into three topological types as ``quasi-$\beta$'', ``writhed'' and ``top-to-top''.
\romannumeral2 ) Among them, the ``writhed'' and ``top-to-top'' types tend to show high flare activity.
\romannumeral3 ) As the signs of twist and writhe agree with each other in most cases of  the ``writhed'' type (12 cases out of 13),
we propose a magnetic model in which the emerging flux regions in a $\beta\gamma\delta$ AR are not separated but united as a single structure below the solar surface.
\romannumeral4 ) Almost all the  ``writhed''-type ARs have downward knotted structures in the mid portion of the magnetic flux tube. This, we believe, is the essential property of $\beta\gamma\delta$ ARs.
\romannumeral5 ) The flare activity of $\beta\gamma\delta$ ARs is highly correlated not only with the sunspot area but also with the magnetic complexity.
\romannumeral6 ) We suggest that there is a possible scaling-law between the flare index and the maximum umbral area.
\end{abstract}
\end{opening}

\section{Introduction}
     \label{S-Introduction} 

A sunspot group with umbrae of opposite polarity within a single common penumbra is called the $\delta$-sunspot group (\opencite{Kunzel60}). 
It is well known that major flares almost always occur in $\delta$-sunspot groups (\opencite{Tanaka75}; \opencite{Zirin87}; \opencite{Sammis00}).
Hence, investigation of the formation and evolution of the $\delta$-configuration is the key in understanding the major flare activities.

\inlinecite{Zirin87} classified the formation of $\delta$-spots in three types as follows: \romannumeral1) Emergence of a single complex active region (AR) formed below the surface, {\it i.e.}, the so-called ``island $\delta$-spot'', \romannumeral2) emergence of large satellite spots near a larger and older sunspot, and \romannumeral3) collision of sunspots of opposite polarity from different dipoles.
Then \inlinecite{Ikhsanov04}, from their morphological study of $\delta$-spots, classified the magnetic topology of collisional interaction between two major emerging flux tubes into three types, namely ``vertical (top-by-top) collision", ``lateral (side-by-side) collision", and ``frontal (foot-by-foot) collision" (see also \opencite{Ikhsanov03}).
Although these researches show a variety of ways of $\delta$-spot formation, they did not give us a detailed physical view of the formation of $\delta$-spots.
This is because their classifications were phenomenological.

Succeeding researches (\opencite{Kurokawa87}; \opencite{Tanaka91}; \opencite{Leka96}; \opencite{vanDriel97}; \opencite{Ishii98}; \opencite{Linton99}; \opencite{Fan99}; \opencite{Lopez00}) have found several physical characteristics of active $\delta$-spots, in that they have strong magnetic shear structures in their magnetic neutral line area and  their opposite polarities rotate around each other.
These characteristics were interpreted as due to the emergence of a kinked flux rope.
\inlinecite{Ishii98} and \inlinecite{Kurokawa02} proposed that the $\delta$-spot is produced by the emergence of a twisted magnetic flux rope whose strong shear produces high flare activities.
\inlinecite{Poisson13} suggested that the $\delta$-spot could be due to the emergence of a deformed single tube which has a downward convex structure at its middle part with kinking.
Thus the twist and writhe of a magnetic rope is now considered as a key factor for the development of $\delta$-spots and the associated flare activities which sometime show sigmoidal morphology (\opencite{Rust96}).

Although a number of case studies revealed the important characteristics of $\delta$-spot evolution, we have not yet had a conclusive observational and physical view of the $\delta$-spot formation.
In this article, we tried to obtain more clues on this issue by surveying the $\delta$-spots in solar activity cycle 23 paying attention to magnetic helicity (twist and writhe).
Among all the $\delta$-spots in Cycle 23, we selected those which can be studied from their birth on the visible solar disk in order to unambiguously study the magnetic connection between component sunspots in a group.
Further we limited the samples to the $\beta\gamma\delta$ regions, as they show the highest flare activity (\opencite{Sammis00}).
The magnetograms and the continuum images taken with the {\it Michelson Doppler Imager} (MDI; \opencite{Scherrer95}) on board the {\it Solar and Heliospheric Observatory} (SOHO; \opencite{Domingo95}) were used to follow the evolution of the ARs.

Our study shows that the flare-active $\delta$-spots are mainly formed by the emergence of writhed and twisted magnetic tubes, which appear as a quadrupolar magnetic configuration on the photosphere.
Even the appearance of a more complex $\delta$-configuration can be interpreted as a modification to this basic configuration.
On the other hand, the flare activity is low when the magnetic tubes in the $\delta$-configuration show insignificant writhe or twist.

In Section 2, we describe the data used in our study and the analysis method.
In Section 3, we present our detailed analysis on some representative cases and summarize our statistical analysis.
Finally, we give discussion on our results and our conclusions in Sections 4 and 5, respectively.

\section{Data and Analysis} 
      \label{S-Data-and-Analysis}   
 \subsection{Region Selection} 
  \label{sec:subsection2-1}  
To select $\beta\gamma\delta$ ARs in Cycle 23, we based our study on the following two data catalogues:
USAF-MWL (also known as USAF-SOON) \footnote{ftp:\slash\slash ftp.ngdc.noaa.gov\slash STP\slash SOLAR\_DATA\slash SUNSPOT\_REGIONS\slash USAF\_MWL} and USAF\slash NOAA sunspot data \footnote{http:\slash\slash solarscience.msfc.nasa.gov\slash greenwch.shtml}.
The former data base is compilation of daily solar reports of six ground-based observatories. The latter mainly consists of data based on space  observations after the launch of the SOHO spacecraft.
When the type of an AR is assigned differently in the two catalogues, we choose the more complex one of the region as the representative.
In this way we selected 200 $\beta\gamma\delta$ regions which are classified so at least once during their lifetime.
Four ARs classified as $\beta\gamma\delta$ in the catalogues were excluded from our dataset as they showed no sign of $\delta$-type in our visual check with SOHO/MDI images.
Finally we found and selected 31 candidates which can be studied from their initial emergence on the visible solar disk with reference to the Solar-Monitor website \footnote{http://www.solarmonitor.org/}.

\subsection{Imaging Data} 
\label{Images} 

To follow the evolutions of ARs, we used the longitudinal magnetograms and the continuum images taken with SOHO/MDI with a cadence of 96 min and a pixel size of $2.0''$.
We utilized the {\it Geostationary Operational Environmental Satellite} (GOES) X-ray data to estimate the total flare activity of selected ARs.
Full-disk Fe {\sc xii} 195{\AA} images of the {\it Extreme ultraviolet Imaging Telescope} (EIT; \opencite{Delabou95}) on board SOHO taken with a cadence of 12 min and a pixel size of $2.6''$ were used to check coronal structures.
The {\it Transition Region and Coronal Explorer} (TRACE; \opencite{Handy99}) images in 195{\AA}, 1600{\AA}, and 5000 {\AA} taken with approximately 1 min cadence and $0.5''$ per pixel were also used if available. 

\subsection{Image Analysis} 
\subsubsection{Alignment of AR Images}
\label{Alignment} 
\begin{figure}[t]
\begin{center}
\includegraphics[scale=0.6,clip,trim=0 0 0 0]{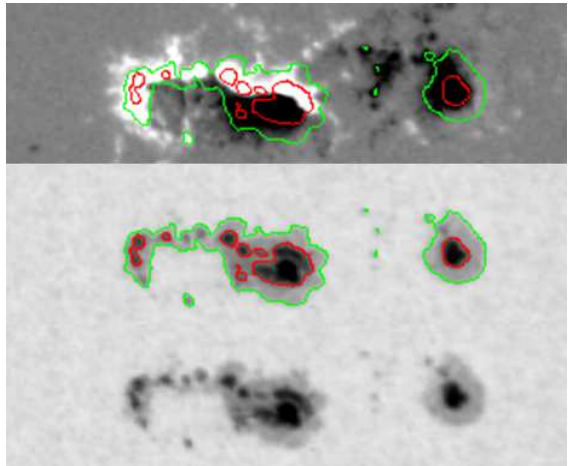}
\caption{
A magnetogram (top, gray scale) and a white light image (bottom) of NOAA 10314.
Green and red contours show the intensity levels of 0.9 and 0.65 of the quiet region (QR) intensity and delineate QR-penumbra boundaries and penumbra-umbra boundaries, respectively (the middle panel).
The AR has $\delta$-configuration in the central part.
}
\label{Sample-image}
\end{center}
\end{figure}

A time-series of MDI images of a target AR was aligned by the method proposed by \inlinecite{Chae01} and \inlinecite{Chae_etal01}. 
We converted original 2 arcsec resolution images into 1 arcsec ones, and then applied a non-linear mapping to remove the differential rotation effect.
With these aligned images, we studied the evolution of the target AR, and measured the duration of the $\delta$-state by comparing the magnetograms with the continuum images.
In this study, we set the brightness levels of 0.9 for penumbrae and 0.65 for umbrae relative to the intensity of the quiet region (see Figure~\ref{Sample-image}).

\subsubsection{Circulation}
\label{Circulation}

When we closely looked at the evolution of an AR, some sunspots exhibited prominent rotation around their centers.
We estimated the magnitude of rotation by measuring the spatially integrated vorticity  around the rotating sunspot in the following manner.
First, we re-aligned the sunspot location in each sequential magnetogram to the center of gravity of the magnetic field strength to remove the effect of proper motion of the sunspot.
Second, we applied the local correlation tracking method(LCT; \opencite{November88}) to these images to derive the horizontal velocity field
 (${\bf v}$) around the sunspot.
Then, we calculate the area integral of the vorticity, namely circulation ($C$), by

\begin{equation}
C = \int (\nabla\times {\bf v})_{z} dS, \qquad (\nabla\times {\bf v})_{z} = \frac{\partial v_{y}} {\partial x} - \frac{\partial v_{x}} {\partial y}.
\end{equation}

\begin{figure}[t] 
\begin{center}
\includegraphics[scale=0.34,angle=0,clip,trim= 0 0 0 10]{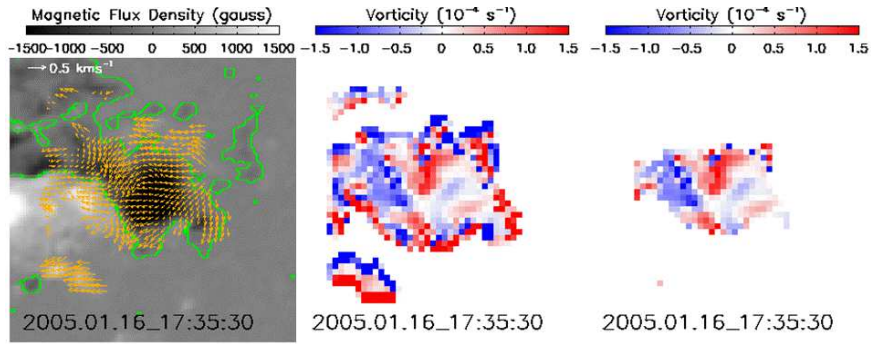}
\caption{Erosion operation applied to the vorticity field at 17:35UT on 16 January 2005.
From left to right are shown a magnetogram, a vorticity field, and the edge-trimmed map of the vorticity field, respectively.
The velocity field (yellow arrows) and neutral lines (green contours) are overlaid on the magnetogram (gray scale).}
\label{EROSION-MDFD}
\end{center}
\end{figure}

We trimmed away the edge zone of three-pixel width from the vorticity field to remove fluctuating and erroneous values before the final estimation of the circulation.
The procedure of trimming is done by first making a binary image from the original vorticity map, and then applying the erosion function ERODE in the IDL (Interactive Data Language) software to the binary image which results in the shrinkage of island areas in the binary image by three pixels, and finally masking the original vorticity map with the eroded binary image.
An example of our method for the vorticity field at 17:35UT on 16 January 2005 is shown in Figure~\ref{EROSION-MDFD}.
From left to right, the velocity field overlaid on the magnetogram, the vorticity field, and trimmed-away map of the vorticity field are shown.
Erroneous vorticity values originate at the border between the real velocity field and the quiet region where the velocity of magnetic features is zero.
The erosion operation can eliminate the contaminations from the real rotational elements around the target sunspot effectively as shown in Figure~\ref{EROSION-MDFD}.

According to the mathematical definition of circulation, positive (negative) circulation corresponds to counter-clockwise (clockwise) rotation.
Note that the signs are opposite to those of the magnetic twist described in Section \ref{Helicity Sign}.

\subsubsection{Tilt Angle of $\delta$-Spots}
\label{Tilt angle} 

\begin{figure}[t] 
\begin{center}
\includegraphics[scale=0.5,clip,trim=10 10 10 10]{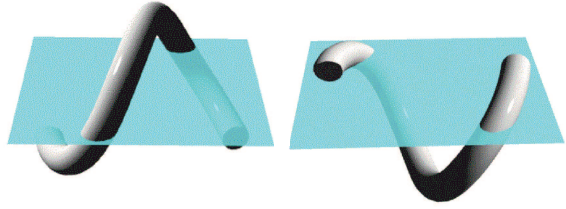}
\caption{Illustrations of the emergence of tubes with left-handed writhe in the cases of upward-kinked and downward-kinked configurations.
In both cases, the inclination angle of the line connecting the two footpoints rotates clockwise and their separation increases or decreases according to the orientation of the kink.}
\label{LHD-INC}
\end{center}
\end{figure}

The temporal variation of the tilt angle ({\it i.e.} the inclination angle of the line connecting the bipoles) tells us the orientation of the kink in an emerging magnetic tube.
Figure~\ref{LHD-INC} schematically shows the case of a tube with left-handed writhe; upward-kinked (left panel) and downward-kinked (right panel).

When they emerge through the photosphere, the tilt angle rotates in the clockwise (CW) direction in both cases.
The distance between the footpoints increases for the upward-kinked case, while it decreases for the downward-kinked case.
When the tubes with right-handed writhe emerge, the rotation is in the counter-clockwise (CCW) direction and the distance between the footpoints changes according to the kink direction as in the left-handed case.
Thus we can infer the orientation of the kink in the emerging magnetic tubes from the temporal variations of the inclination angle and the distance between the two spots. 

\subsubsection{Sign of Magnetic Helicity}
\label{Helicity Sign}

\begin{figure}[t] 
\begin{center}
\includegraphics[scale=0.45,clip]{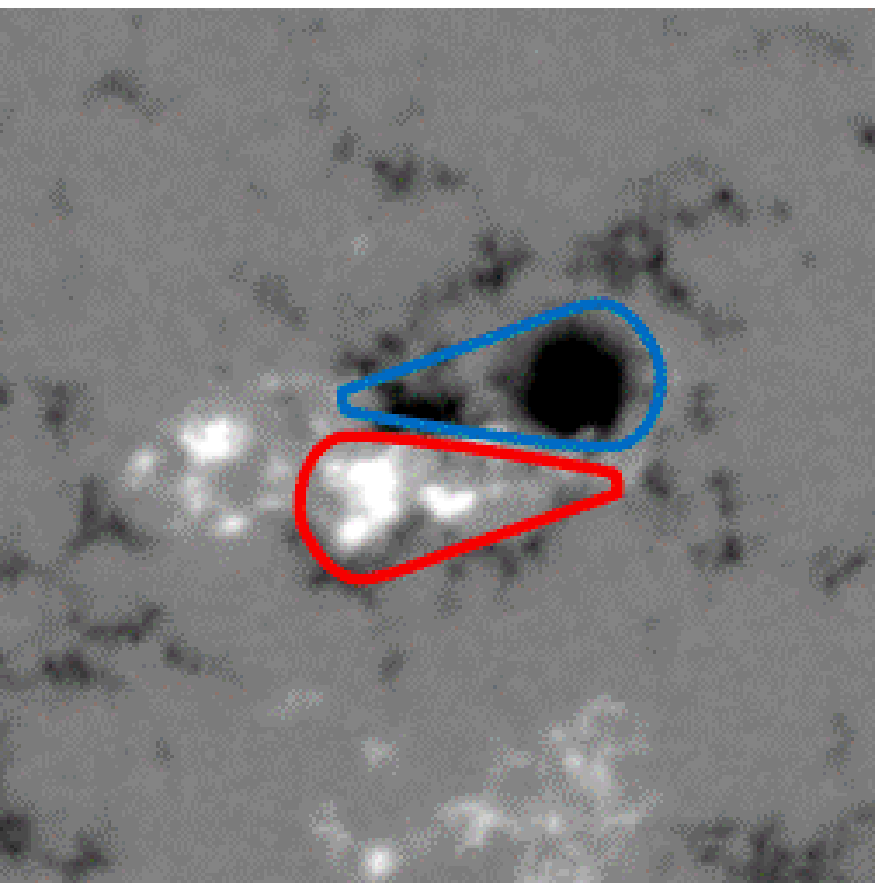}
\hspace{3mm}
\includegraphics[scale=0.45,clip]{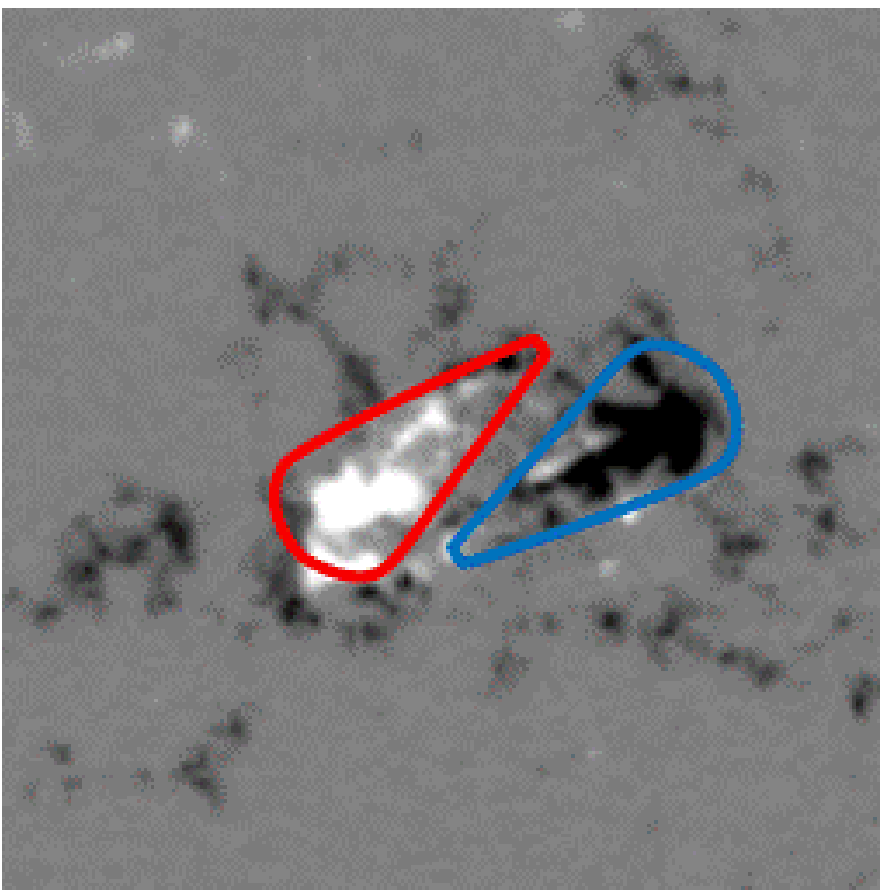}
\caption{Sample images of magnetic tongues.
When the head parts of the ``Yin-Yang" structure forming the magnetic tongues show clockwise or counter-clockwise orientation, they indicate negative or positive magnetic helicity, respectively.}
\label{Tongue}
\end{center}
\end{figure}

 A ``magnetic tongue" pattern is proposed as a useful parameter to give the sign of magnetic helicity of the emerging field (\opencite{Lopez00}; \opencite{Lopez08}; \opencite{Luoni11}). 
When a twisted tube emerges above the photosphere, the spatial distribution of photospheric magnetic field has an elongated shape like a tongue.
The tongues of different polarities show anti-parallel orientations along the neutral line.
Since the orientation of a tongue pattern depends on the sense of twist of the emerged tube, we can use this parameter as a proxy of the sign of helicity (twist) of the emerging magnetic tube as shown in Figure~\ref{Tongue} (see also Figure 1 of \opencite{Luoni11}).
The circulation of a sunspot is another measure of the sign of helicity of the magnetic flux tube, and consistency can be checked by comparing of these two signs.

To avoid confusion, we here reiterate the  definitions for the handedness; the sign of the left (right)-handed twist or writhe is negative (positive).

\subsubsection{Flare Index}
\label{FI}

Several authors have used the soft X-ray flare index (FI) as a proxy of the activity level of ARs, as given by  ({\it e.g.} \opencite{Antalova96}; \opencite{Joshi04}; \opencite{Abramenko05}; \opencite{Jing06})
\begin{equation}
{\rm FI} = 1.0 \times \sum_{i} m_{\rm C} + 10.0 \times \sum_{j} m_{\rm M} + 100.0 \times \sum_{k} m_{\rm X},
\end{equation}
where $m_{\rm C}$, $m_{\rm M},$ and $m_{\rm X}$ are the GOES soft X-ray peak intensity magnitudes (from 1.0 to 9.9) of C, M and X-class flares.
The indices {\it i}, {\it j}, and {\it k} designate flares in each class per unit time period.
For an AR, two kinds of periods were considered in this article; those in which the AR was visible on the disk and the AR was in the $\delta$-state.
We do not consider small flares less than the C-class in this study because they tend to be frequently hidden by major flares or are indistinguishable from the background level.

\section{Results}
\label{S-results} 

The $\delta$-spots studied in this article can be classified, from the point of view of the emergence morphology, the magnetic connection, the helicity of sunspots, and the flare activity, into the following groups: \romannumeral1) top-to-top collision, 
\romannumeral2) writhed loop,
\romannumeral3) quasi-$\beta$, and 
\romannumeral4) the others.
The majority of $\delta$-spots in our sample are classified into the first three groups.
In the next subsection, we select and describe the features of a representative example of each group to show our classification scheme.

 \subsection{Case Studies}
\label{Cases} 

\subsubsection{Evolution of NOAA10720}

NOAA10720 was the most flare-productive region in our samples.
It had the longest $\delta$-duration (longer than 213 h), and showed rapid expansion since its emergence.

\begin{figure}[H] 
\begin{center}
\includegraphics[scale=0.42,angle=0,trim= 20 5 0 10]{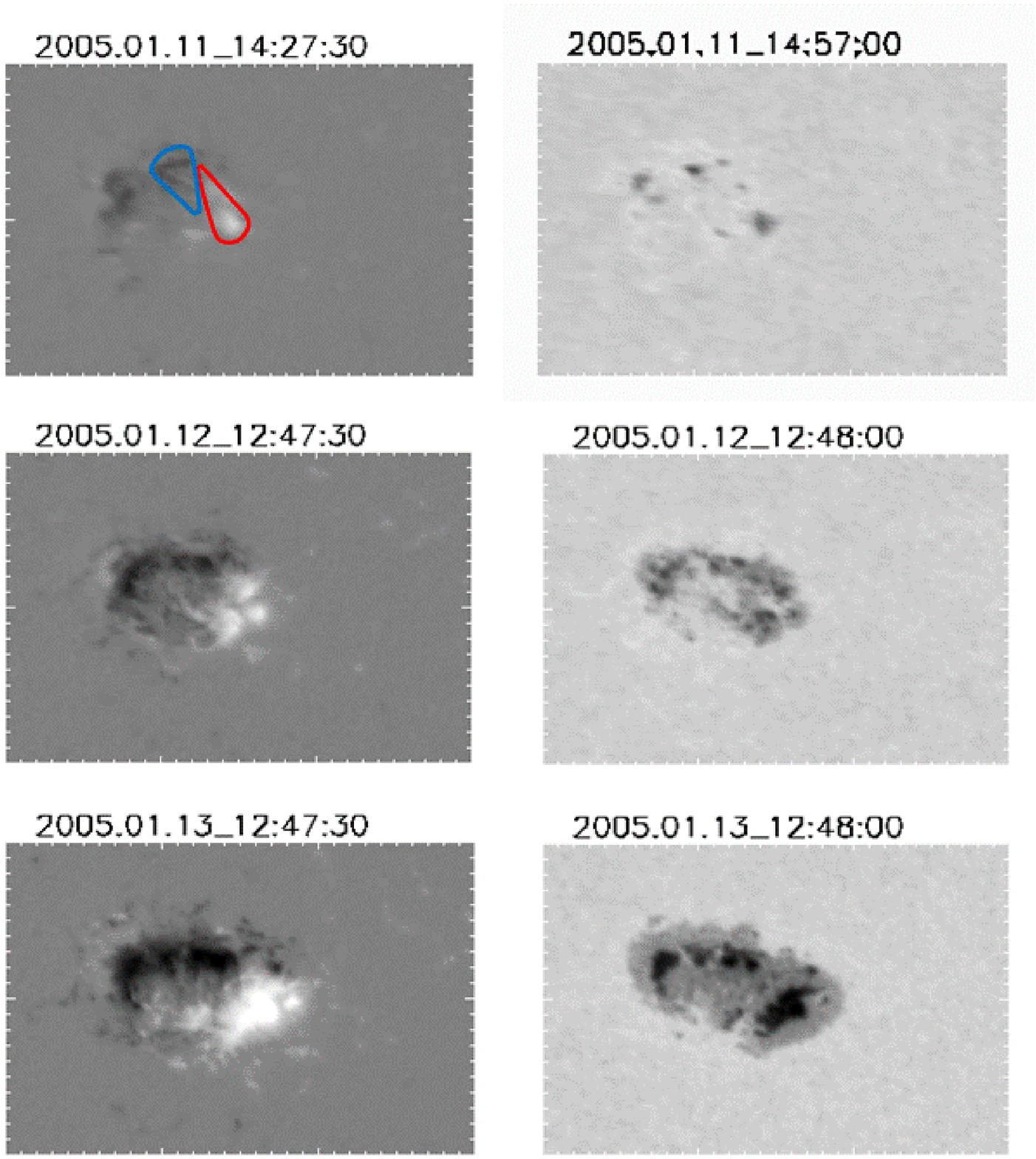}  
\includegraphics[scale=0.42,angle=0,trim= 0 0 0 15]{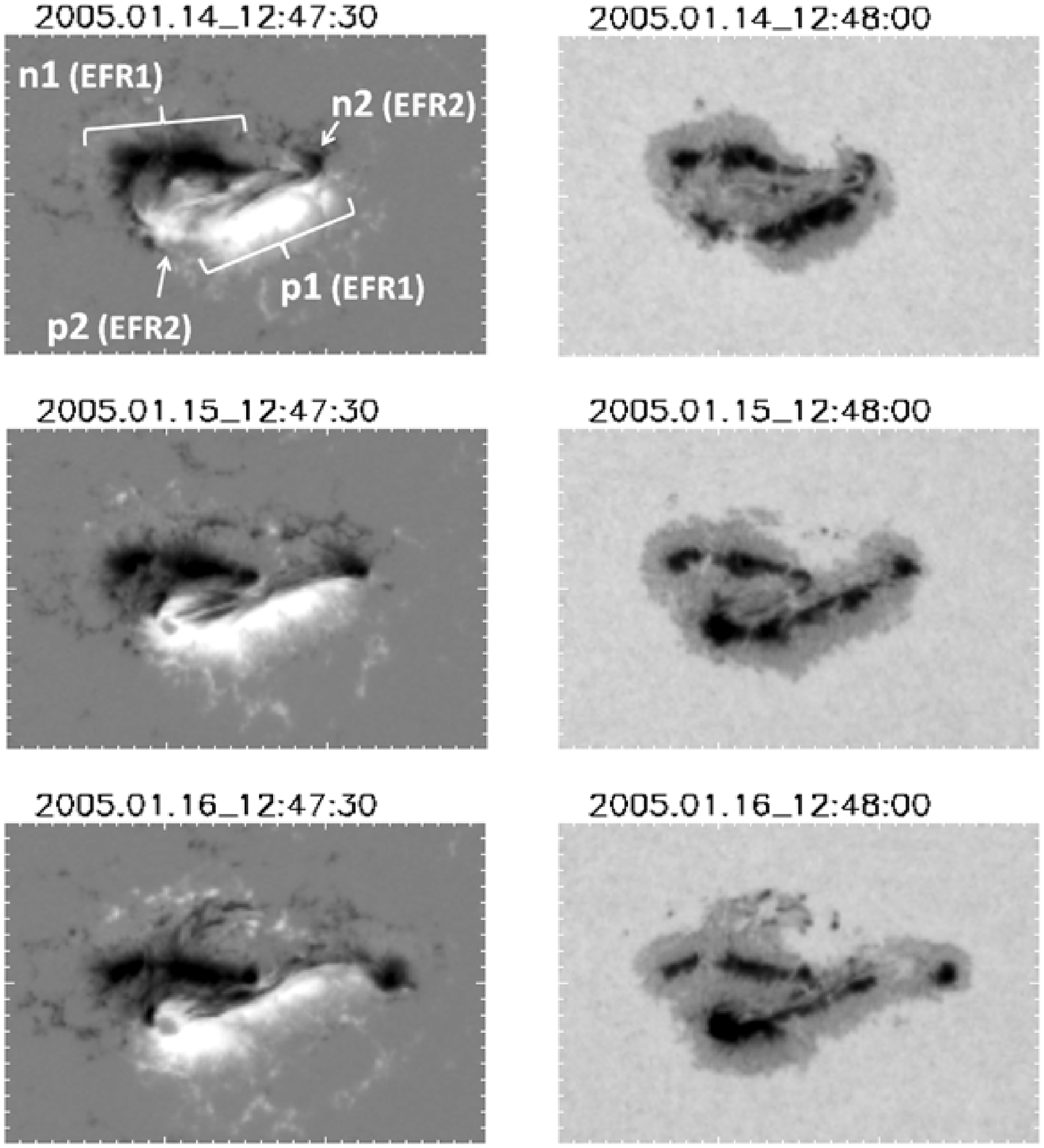}
\caption{SOHO/MDI magnetograms (left) and white-light images (right) of NOAA 10720 corresponding to subsequent days of observations. 
Identified two EFRs are shown in the fourth row. 
The field of view (FOV) of each panel is 300 $\times$ 200 arcsec.}
\label{10720}
\end{center}
\end{figure}

The magnetic and morphological evolution of the region was as follows.
First, an emerging flux region (EFR1) emerged as a simple $\beta$ magnetic configuration with almost east-west orientation obeying Hale's law.
Then it expanded rapidly in size and evolved to the $\delta$-configuration on 12 January 2005 as shown in Figure~\ref{10720}.
Another new highly sheared EFR (EFR2), which disobeyed Hale's law, appeared on the neutral line as if it tried to penetrate into the opposite polarity portion of EFR1 at around 16:00UT on 13 January 2005.

In Figure~\ref{10720}, we cannot see any signature of writhe in EFR1, as its tilt showed no systematic rotational movements.
The magnetic tongues of EFR1 in the initial emerging phase indicated negative magnetic helicity as shown in the first row of Figure~\ref{10720}.
This means that the tube emerged with a left-handed twist.
Unfortunately, the tongue pattern could not clearly be seen after the vigorous and complex development of EFR1.

On the other hand, the tilt angle of EFR2 showed a slow CW rotation, suggesting the emergence of a loop with left-handed writhe.
The twist of the magnetic tubes can be estimated by the circulation around the footpoints of the magnetic tubes, {\it i.e.} sunspots.

\begin{figure}[H] 
\begin{center}
\includegraphics[scale=0.35,angle=0,clip,trim= 0 0 0 0]{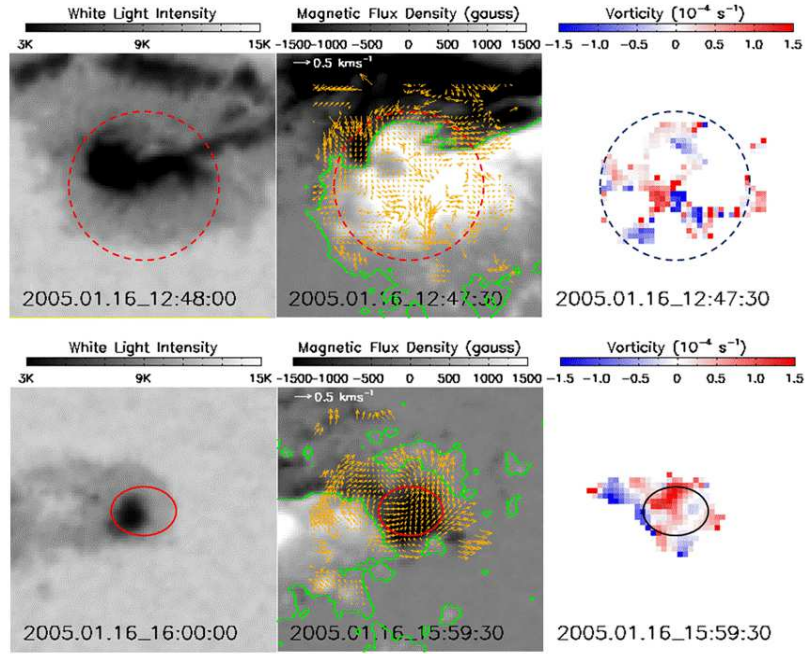}
\caption{
White light images (left), magnetograms (middle), and vorticity fields (right) of the positive (top row) and negative (bottom row) polarity areas of EFR2, respectively.
The size of each FOV is 100 $\times$ 100 arcsec, and its center is at the center of gravity of magnetic flux distribution. In the middle panels are overlaid the velocity fields (yellow arrows) and the neutral lines (green contours).
The areas used to compute the circulation are shown by a circle of 28 arcsec radius in the top row and by an ellipse in the bottom row whose major and minor axes are 24 and 18 arcsec, respectively. 
}  
\label{10720-Vor-map}
\end{center}
\end{figure}

\begin{figure}[H] 
\begin{center}
\includegraphics[scale=0.22,angle=0,clip,trim= 30 100 50 25]{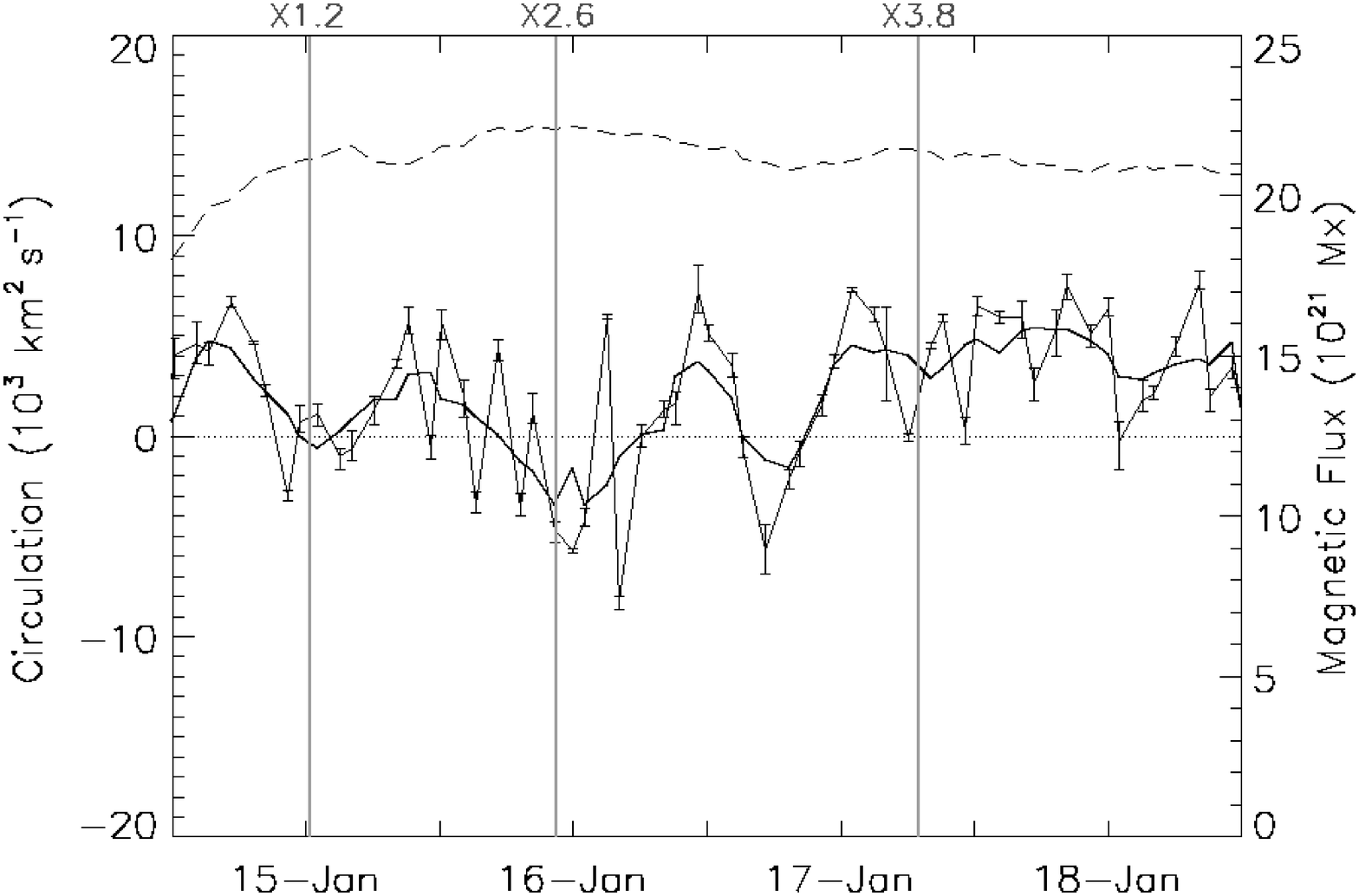}
\includegraphics[scale=0.22,angle=0,clip,trim= 0 101 50 28]{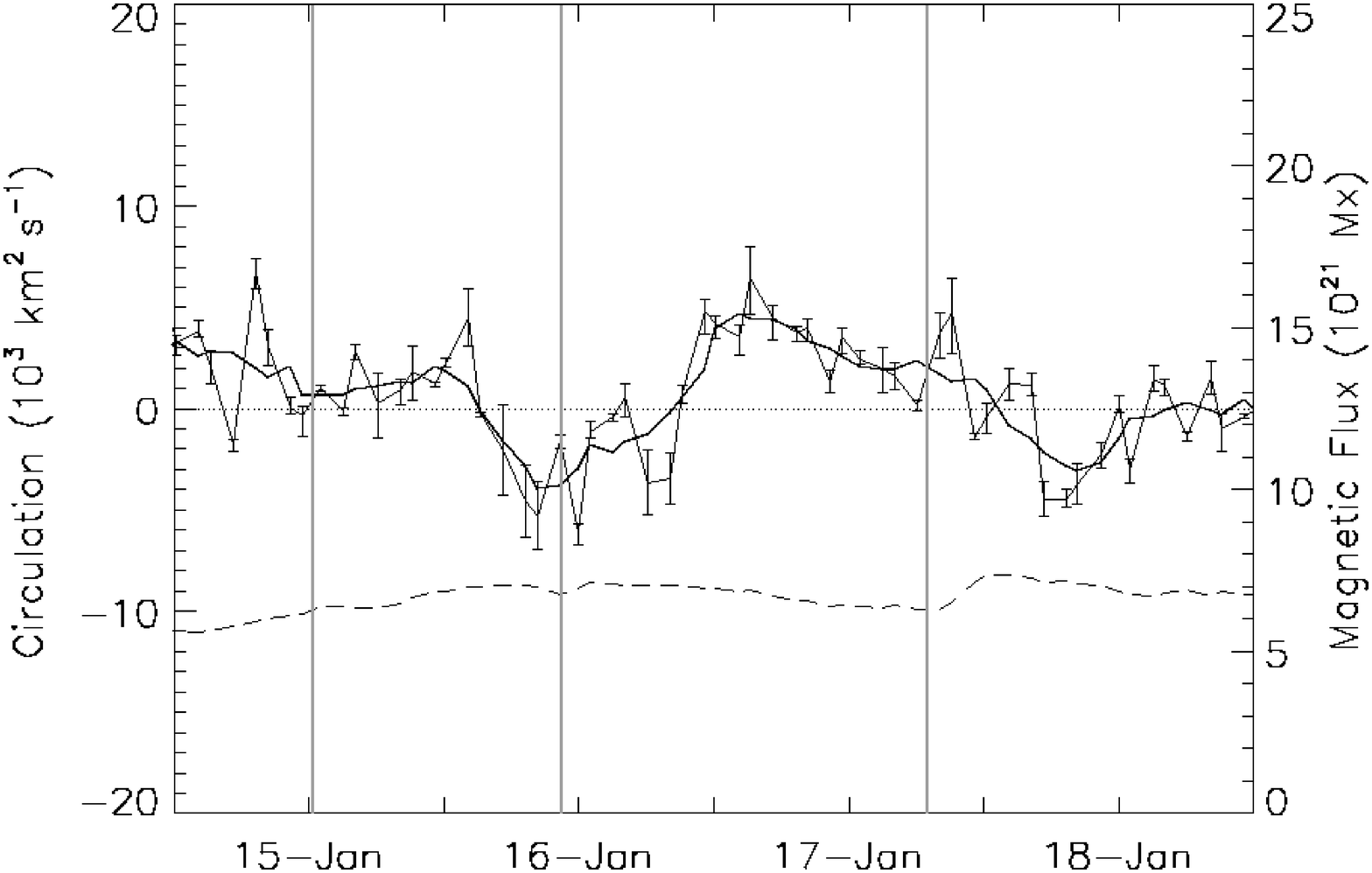}
\caption{
Time variations of circulation (thin solid lines) and magnetic flux (dashed lines) in sunspots p2 (top) and n2 (bottom) in NOAA 10720.
The thick solid line shows the running average of circulation over five consecutive data points in each panel.
The error bars indicate a possible range of the circulation values when the size of the analysis area is changes (see text).
Vertical lines represent the times of the X-class flares. The magnetic flux values are evaluated for the full FOV of Figure~\ref{10720-Vor-map}.
}
\label{10720-Vor-plot}
\end{center}
\end{figure}

Now let us see the distribution of vorticity around sunspots of p2 and n2 shown in Figure~\ref{10720-Vor-map}.
By specifying the integration area with a circle or an ellipse, which covers the target sunspot as compactly as possible, we estimated the circulation over the area covering each sunspot.
The reliability of the circulation value was estimated by changing the size of the calculation area.
The area was enlarged or reduced by 2 arcsec for this purpose.
Besides the procedures described in Section~\ref{Circulation}, we masked the opposite polarity areas before the integration to avoid contamination. 
The temporal evolution of the circulation of each sunspot is displayed in Figure~\ref{10720-Vor-plot}.
Both sunspots showed generally positive circulation although they showed negative values at times.
When we closely looked at the vorticity map and the motions of small sunspots in the neighborhood of the n2 spot for a day on 16 January, we found that the rapid sliding motions of small sunspots relative to the static environment were recorded as false negative circulation.
If we take this contamination into account, we can assume that the circulations around the p2 and n2 spots were mostly positive during the evolution, which means that the footpoints of the magnetic tube of EFR2 showed a CCW rotation during their evolution.
Therefore, EFR2 can be considered as an emerging loop of a left-handed twist.

\begin{figure}[H] 
\begin{center}
\includegraphics[scale=0.32,angle=0,clip,trim= 0 15 0 46]{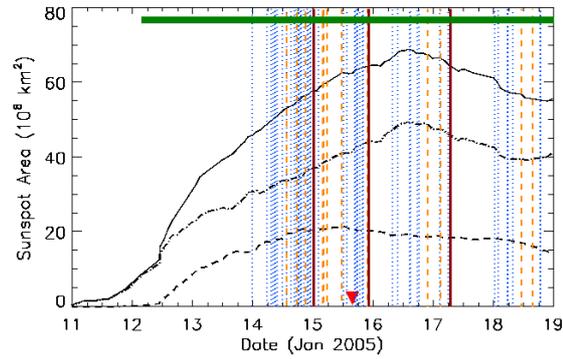}
\caption{
Time evolution of umbra (dashed), penumbra (dash-dotted), and total sunspot (solid) areas in NOAA 10720.
Vertical lines indicate X-class (brown solid), M-class (orange dashed), and C-class (blue dotted) flares, respectively.
The horizontal thick green line at the top represents the duration of the $\delta$-state.
The triangle at the bottom indicates the central meridian passage of the region.
}
 \label{10720-Area-Flare}
\end{center}
\end{figure}

\begin{figure}[H] 
\begin{center}
\includegraphics[scale=0.45,angle=0,clip,trim= 0 0 0 0]{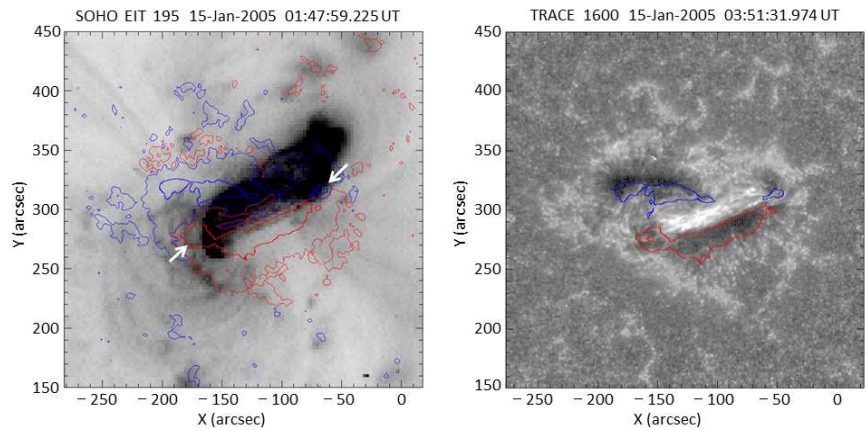}
\caption{
Left: A SOHO/EIT 195{\AA} image (in negative print) of NOAA 10720 taken at 01:47 UT, 15 January 2005, after the X1.2 flare peaked at 01:02 UT.
White arrows indicate two polarities of EFR2. An inverse-S shaped structure is clearly seen. 
Right: A TRACE 1600{\AA} image taken at 03:51 UT on 15 January 2005 between the two X-class flares.
A bright structure existed between the two polarities of EFR2 even in a relatively quiet period of flare activity.
Overlaid contours (red for positive and blue for negative polarities) show the line-of-sight magnetic fields of 1500 gauss (G) and 100 G (only in the left panel).
The FOV of both images is 300 $\times$ 300 arcsec.
}    
\label{10720-EUV-image}
\end{center}
\end{figure}

The timing of the flaring in the region is shown in Figure~\ref{10720-Area-Flare}.
After the emergence of EFR2 on 13 January, many flares have occurred, including three X-class flares.
High activity triggered by the emergence of EFR2 reminds us of the top-to-top collision model proposed by \inlinecite{Ikhsanov04}.
The flare morphology gave us additional information on the magnetic helicity of EFR2.
The SOHO/EIT 195 {\AA} image in the left panel of Figure~\ref{10720-EUV-image} shows an inverse-S shaped structure over EFR2 after the X1.2 flare that peaked at 0:43UT.
The inverse-S sigmoid morphology indicates the left-handed twist of the magnetic tube.
The TRACE 1600 {\AA} image in the right panel of Figure~\ref{10720-EUV-image} shows bright  loops connecting the two polarities of EFR2.
\inlinecite{Guo13} argued that negative helicity was accumulated in EFR2 by a model of non-linear force-free fields, suggesting the emergence of a tube with a left-handed twist.

Thus, we conclude that the significant flare activity of NOAA 10720 is due to the emergence and the intrusion of the twisted and writhed EFR2 into the magnetic neutral line of the pre-emerged EFR1.
 This region is a typical example of the top-to-top collision formation of an active $\delta$-type spot. 

\clearpage

\subsubsection{Evolution of NOAA10826}
\label{S-10826} 

A day after the emergence of 29 November 2005, NOAA10826 formed a $\delta$-configuration and showed high flare activity in the $\delta$-duration of 56 h.

\begin{figure}[H] 
\begin{center}
\includegraphics[scale=0.42,clip,trim=  45 0 0 10]{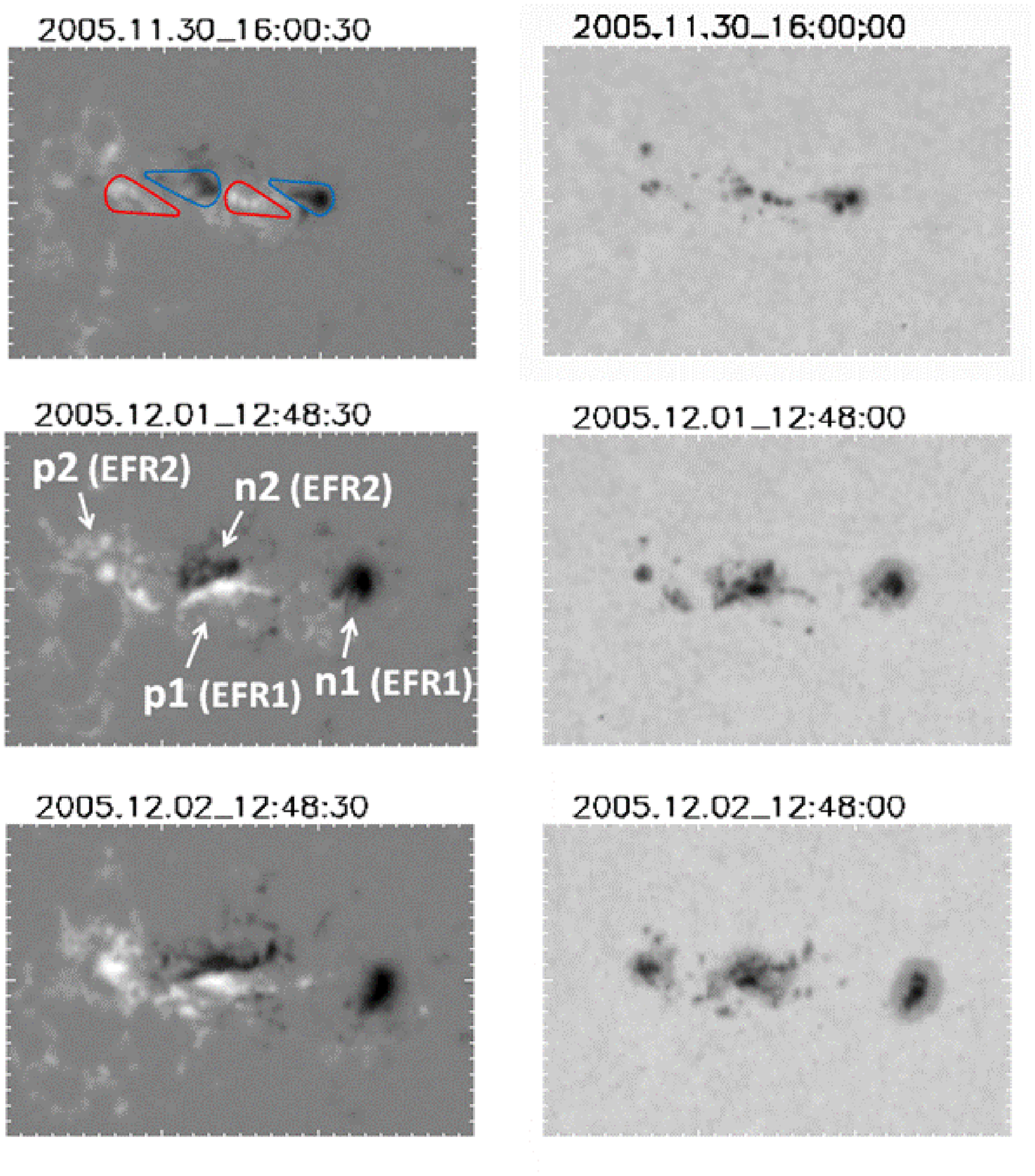} 
\includegraphics[scale=0.42,clip,trim=   0 0 0 15]{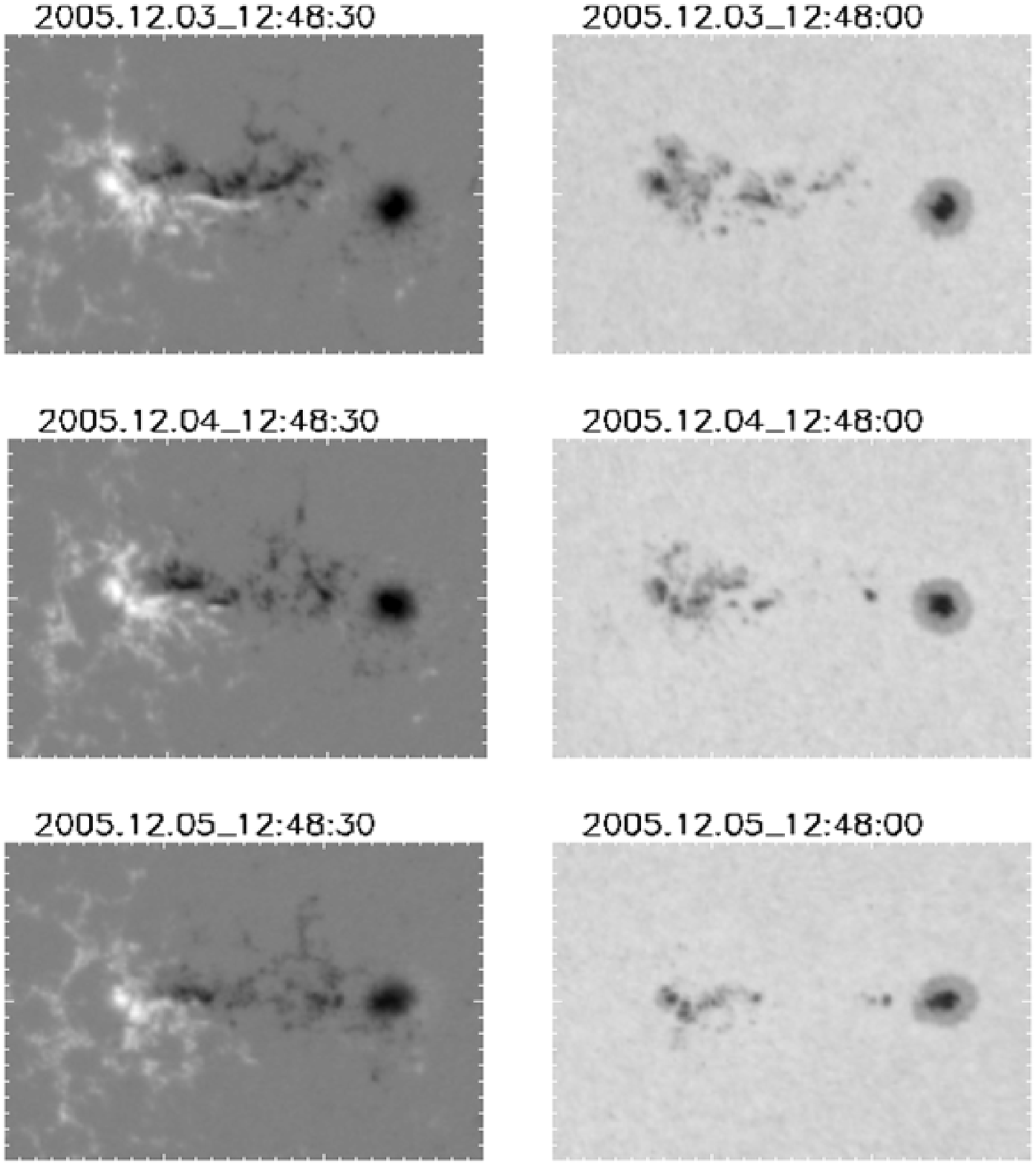}
\caption{
SOHO/MDI magnetograms (left) and white-light images (right) of NOAA 10826. 
Identified two EFRs are shown in the second row. 
The FOV of each panel is 300 $\times$ 200 arcsec.
}
\label{10826}
\end{center}
\end{figure}

Its magnetic and morphological evolution is shown in Figure~\ref{10826}.
Two EFRs emerged in succession along the east-west line.
The $\delta$-configuration was formed between the following sunspot of EFR1 and the preceding sunspot of EFR2.

The magnetic tongue patterns of EFR1 and EFR2 showed the sign of negative helicity (the top row of Figure~\ref{10826}), suggesting the emergence of magnetic tubes with a left-handed twist.
The tilt of the $\delta$-configuration rotated in the CW direction, with an average rotational speed of $35^\circ$ day$^{-1}$, which is similar to the speed of $30^\circ$ day$^{-1}$ measured in NOAA10314 by \inlinecite{Poisson13}.
As the separation between the umbrae of the $\delta$-spot decreased in the first stage, this $\delta$-configuration behaved as the emergence of a tube with left-handed writhe and downward kink.
As the twist had the same sign as the writhe, we suggest that the tube was kinked downward due to transformation into the writhe from the twist in the layers below the photosphere.

\begin{figure}[H] 
\begin{center}
\includegraphics[scale=0.22,clip,trim=30 100 50 25]{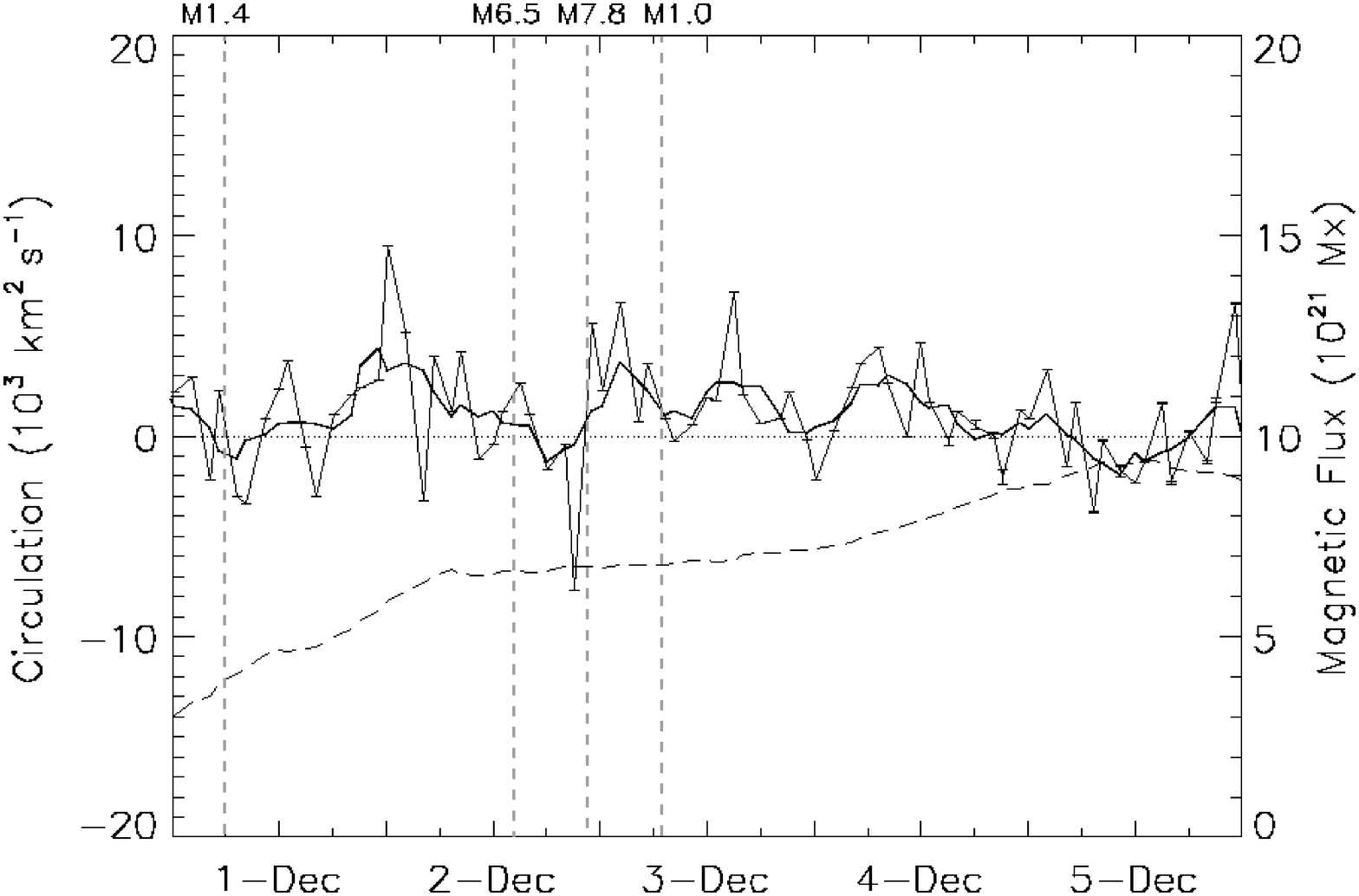}
\includegraphics[scale=0.22,angle=0,clip,trim= 0 101 50 28]{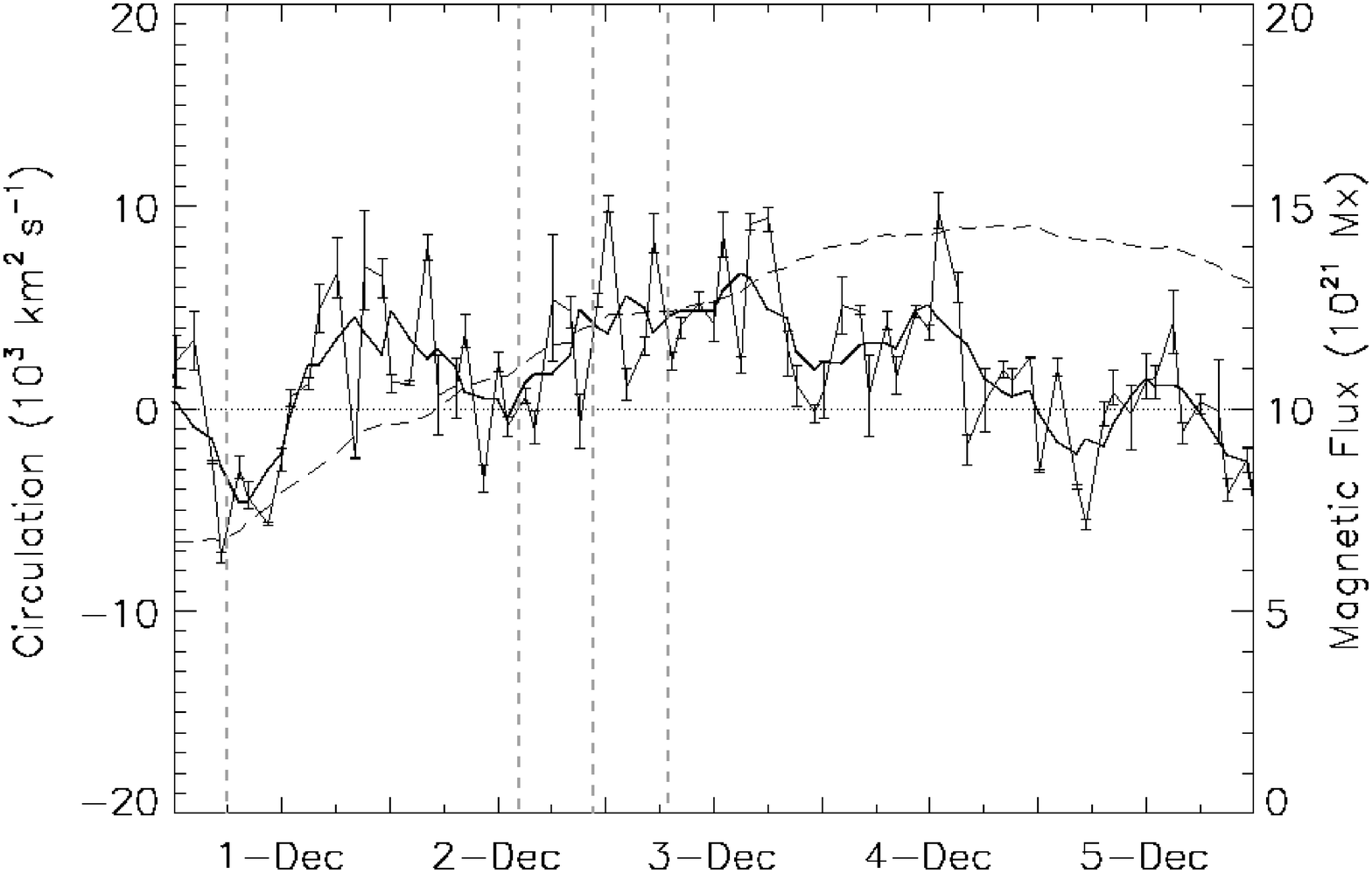}
\caption{
Time variations of circulation (thin solid lines) and magnetic flux (dashed lines) in sunspots n1 (top) and p2 (bottom) in NOAA 10826.
The format is the same as in Figure~\ref{10720-Vor-plot}.
}
\label{10826-Vor}
\end{center}
\end{figure}

The magnetic twist of EFR1 and EFR2 was studied by the circulation at the footpoints of EFRs.
We have selected sunspots n1 of EFR1 and p2 of EFR2 for the evaluation of circulation, as they were isolated and free from possible contamination.
From the temporal variations of the circulation shown in Figure~\ref{10826-Vor}, we conclude that the circulations in both sunspots were predominantly positive, which means that the sunspots in the two EFRs rotated in the CCW direction, {\it i.e.} the tubes emerged with a left-handed twist.

\begin{figure}[H] 
\begin{center}
\includegraphics[scale=0.32,clip,trim=30 5 30 5]{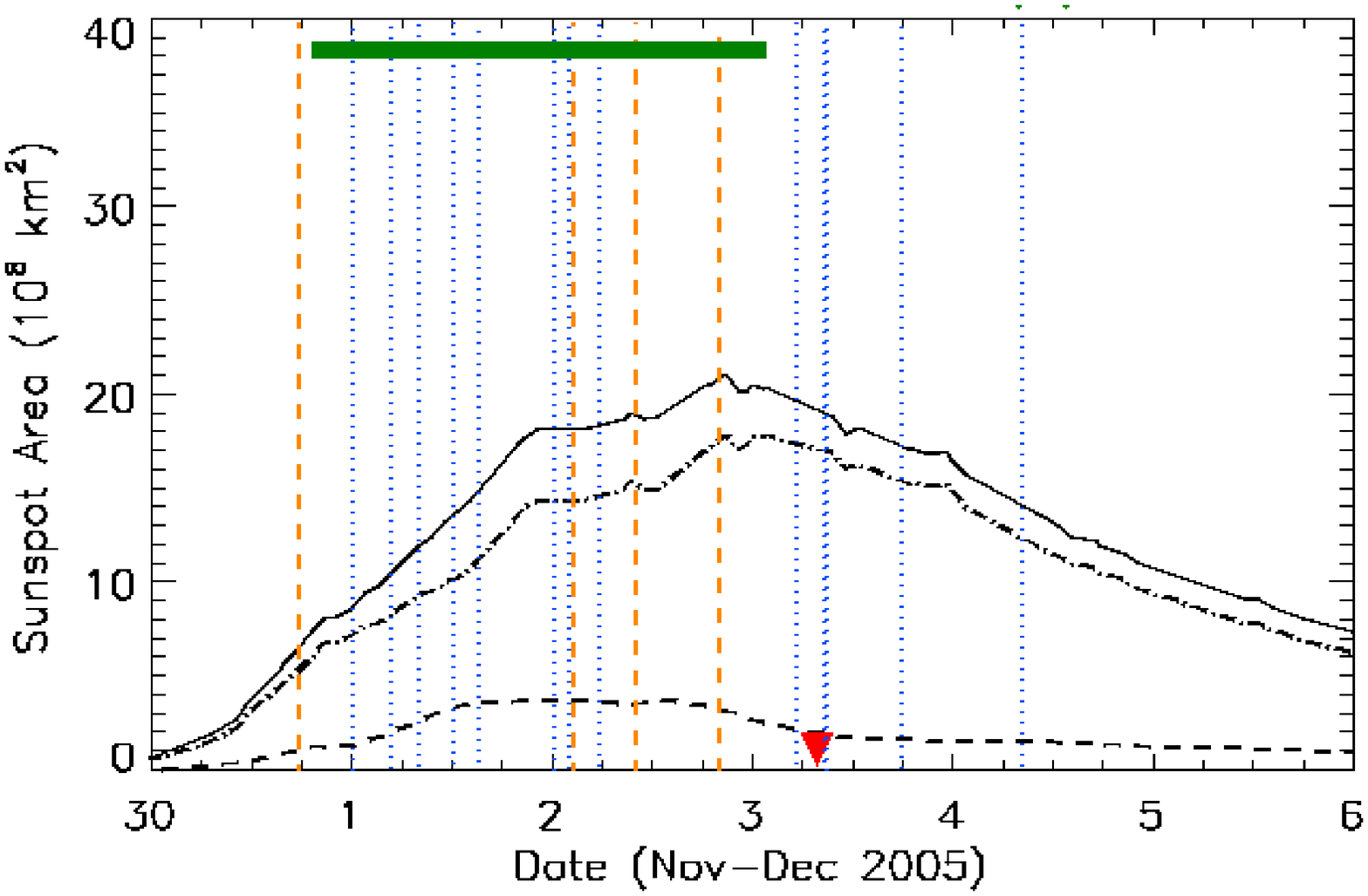}
\caption{
Time evolution of umbra (dashed), penumbra (dash-dotted), and total sunspot (solid) areas in NOAA 10826.
The format is the same as in Figure~\ref{10720-Area-Flare}.
}
\label{10826-Area-Flare}
\end{center}
\end{figure}

\begin{figure}[H] 
\begin{center}
\includegraphics[scale=0.36,clip,trim=10 7 10 10]{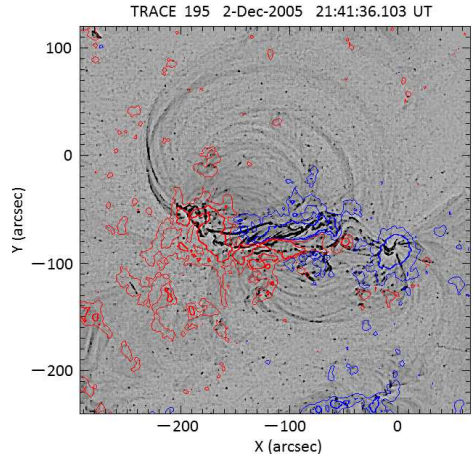}
\caption{
A high-pass filtered TRACE 195 {\AA} image (in negative print) of NOAA 10826.
Contours indicate line-of-sight magnetic fields of 500 G (thick) and 100 G (thin) in positive (red) and negative (blue) polarity areas.
After the M1.0 flare, highly-sheared bright structures appeared on the $\delta$-part of the region.
}
\label{10826-EUV}
\end{center}
\end{figure}

High flare activity in this region started just after the formation of the $\delta$-configuration, producing 17 flares in total (4 M-class and 13 C-class) during its disk passage and 11 flares (3 M-class and 8 C-class) during the $\delta$-configuration (Figure~\ref{10826-Area-Flare}).
The flare morphology gave some insight on the topology near the neutral line of the $\delta$-configuration.
Figure~\ref{10826-EUV} shows the case of an M1.0 flare.
We can see that the bright flare strand had a highly sheared structure with a left-handed twist.
Since the bright strand was probably produced by the magnetic reconnection in the corona, we can infer that the magnetic field near the neutral line had left-handed helicity before the flare explosion.

All the direct or indirect evidence indicates that the EFRs in this region had the same left-handed helicity.
As the magnetic fluxes of sunspots n1 and p2 were of the same order of magnitude (Figure~\ref{10826-Vor}), it is natural to assume that the two EFRs were not physically separated but connected like a single writhed structure as in the case of NOAA 10314 reported by \inlinecite{Poisson13}.
Therefore we classify this region as the ``downward knotted" type.

\subsubsection{Evolution of AR NOAA10050}
\label{S-10050} 

\begin{figure}[H] 
\begin{center}
\includegraphics[scale=0.55,angle=0,clip,trim= 0 0 0 10]{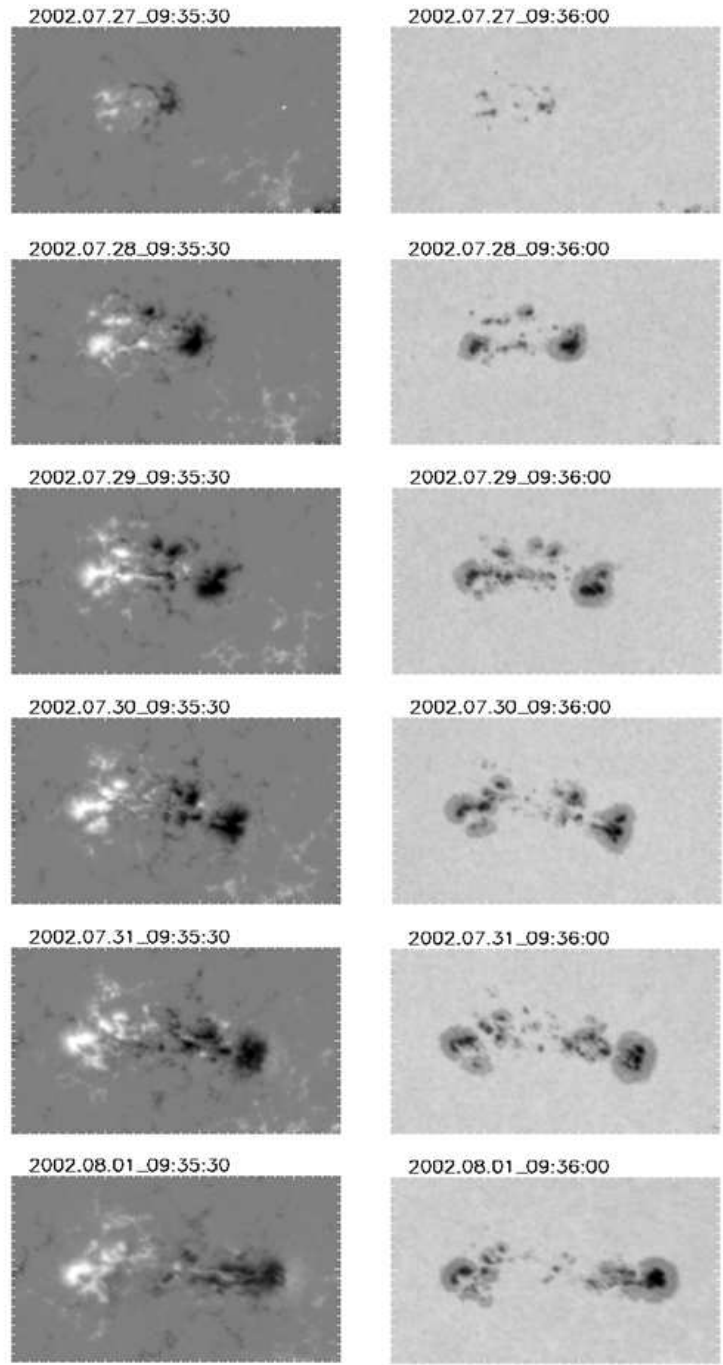} 
\caption{
SOHO/MDI magnetograms (left) and white-light images (right) of NOAA 10050.
The FOV of each panel is 350 $\times$ 200 arcsec.
}
\label{10050}
\end{center}
\end{figure}

NOAA10050 emerged on 26 July 2002.
Although the region developed due to successive emergence of several EFRs, the flare activity was not high.
A few $\delta$-configurations formed and disappeared within their lifetime of less than a day at different locations in the region.
See Figure~\ref{10050} for the magnetic and morphological evolution of the region.

The EFRs in the region emerged with their axes in the east-west direction.
As they evolved, the two opposite polarities of each EFR were separated with each other and sunspots of each polarity merged into a dominant center.
The evolution was very similar to the bipolar emergence model of \inlinecite{Zwaan78}.
The $\delta$-configuration formed transiently when a pair of small opposite polarity elements incidentally came together.
We found no signature of writhe pattern in their motion.

\begin{figure}[H] 
\begin{center}
\includegraphics[scale=0.22,angle=0,clip,trim= 30 100 50 25]{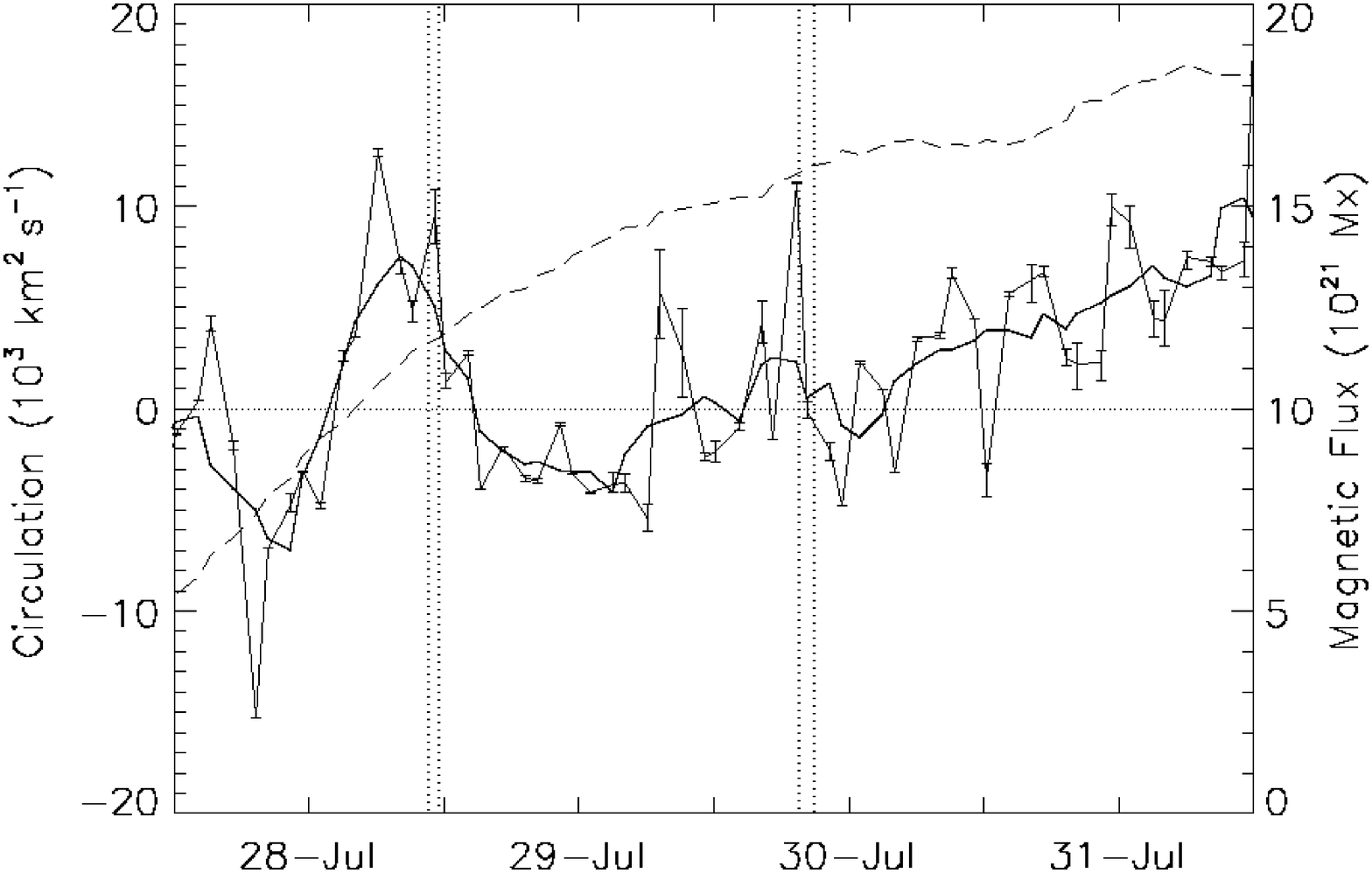}
\includegraphics[scale=0.22,angle=0,clip,trim= 0 101 50 28]{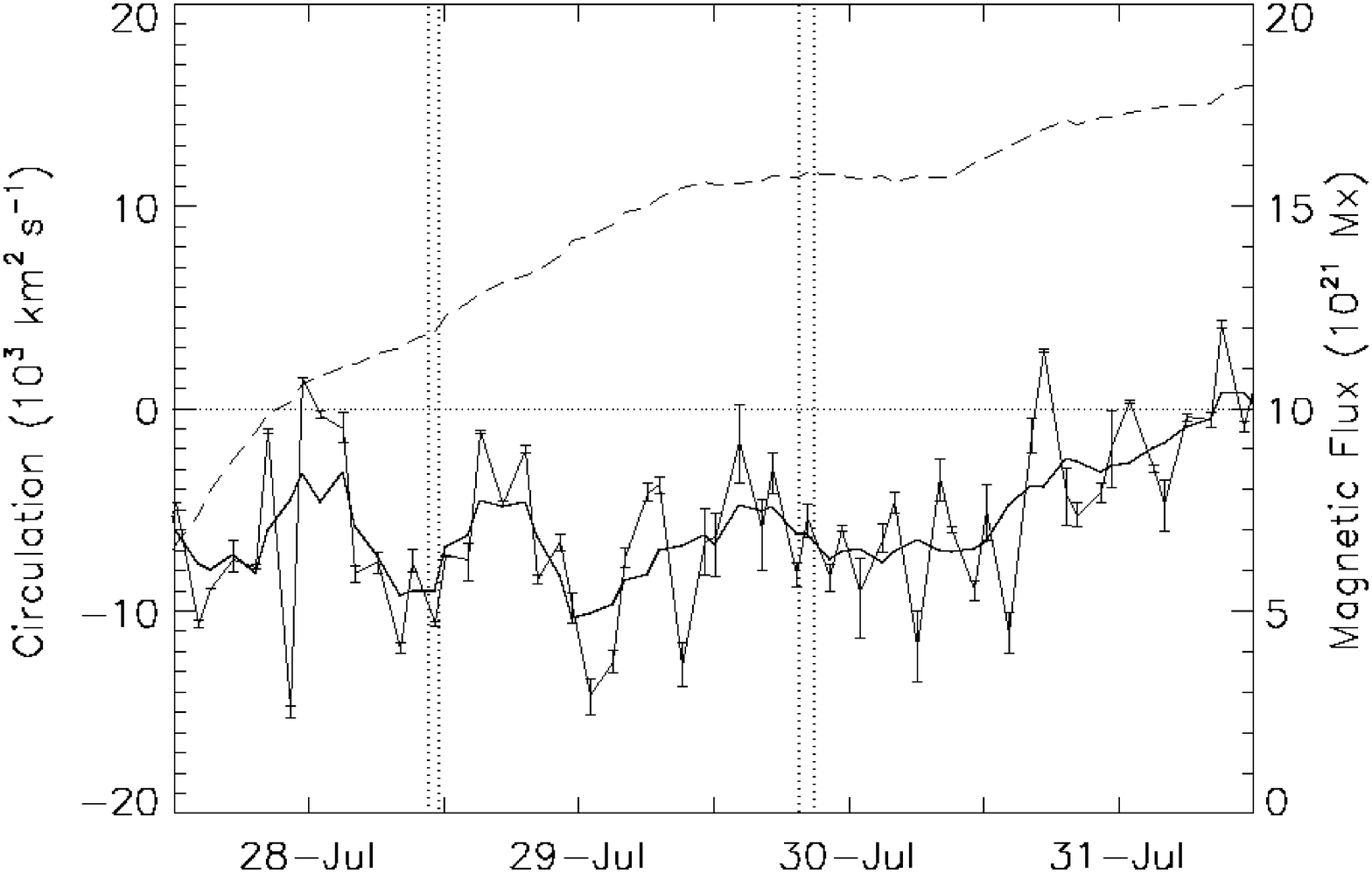}
\caption{
Time variations of circulation (thin solid lines) and magnetic flux (dashed lines) in the preceding negative polality sunspot (top) and the following positive polarity sunspot (bottom) in NOAA 10050.
The format is the same as in Figure~\ref{10826-Vor}.
Dotted vertical lines represent C-class flares.
}
\label{0050_rot}
\end{center}
\end{figure}

\begin{figure}[H] 
\begin{center}
\includegraphics[scale=0.32,angle=0,clip,trim= 0 45 0 46]{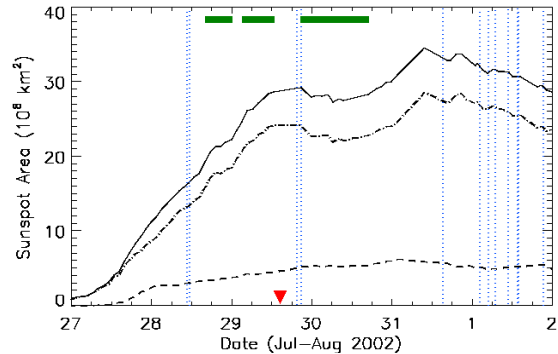}
\caption{
Time evolution of umbra (dashed), penumbra (dash-dotted), and total sunspot (solid) areas in NOAA 10050.
The format is the same as in Figure~\ref{10720-Area-Flare}.
}
\label{0050_area}
\end{center}
\end{figure}

The temporal evolutions of circulation around the dominant polarities are shown in Figure~\ref{0050_rot}.
The preceding negative polarity changed its circulation from negative to positive significantly in the early phase of the evolution.
After several changes in its sign, the positive circulation became dominant in the later phase.
On the other hand, the following positive polarity always indicated negative circulation during the period.
The circulations of the two polarities had a tendency to show opposite directions.
Furthermore, the evolution of circulation was influenced strongly by the successive merging of small sunspots.
Therefore, we may conclude that the actual twist of the dominant sunspots was not high.
The flare activity in the $\delta$-state of this region was very modest as shown in Figure~\ref{0050_area}.

The evolution of the region is very similar to the $\beta$-type and we classify the region as the ``quasi-$\beta$'' type. 

\subsection{Statistical Study}
\label{Statistics}
In our case studies of $\delta$-spots, we notice high flare activity in magnetically complex $\delta$-spots.
In this subsection, we will statistically analyze the dependence of the flare activity on the characteristic parameters of $\delta$-spot regions.

\subsubsection{Parameters}
\label{Parameters}
In Table 1, we summarize several flare indices and the parameters which characterize the 31 ARs we studied.
The peak soft X-ray flux (XR$_{\rm max}$) and FI are the flare indices, while the maximum sunspot area ($S_{\rm max}$), the maximum umbral area  ($S_{\rm Umax}$), $\delta$-term (the duration of the $\delta$-state), the handedness of twist and writhe, and the type of $\delta$-spot formation are the characteristic parameters of ARs.
We will comment on the values of the listed parameters in the following paragraphs.

An AR develops and changes its magnetic type and sometimes passes through the west limb keeping its high activity.
To eliminate possible ambiguity and uncertainty in the estimated parameters, we only considered the events that occurred in longitudes less than W70$^\circ$.
The values of FI were thus counted and listed in Table 1.
The values in parentheses are for the flares that occurred during the $\delta$-state of the region.
The identification of the $\delta$-state was also limited to longitudes less than W70$^\circ$, so that it gives a lower limit in some cases.

The $\delta$-state in an AR was not always continuous in time.
If there appeared multiple $\delta$-states intermittently in an AR, we summed all the time spans of  $\delta$-periods to obtain the ``$\delta$-duration"(or ``$\delta$-term") in this study.
When there was a gap longer than several hours in the available MDI continuum images, we also used TRACE white light images to estimate the $\delta$-duration.
However, there remained the uncertainties ranging from seven to 31 hours, which is regarded as estimation errors for the $\delta$-duration.

As seen in our case studies, the magnetic field distributions in $\beta\gamma\delta$ ARs are complex and the estimation of the magnetic twist from the circulation is sometimes ambiguous.
However, it is well known that the preceding sunspots have more rigid structure and longer lifetime than the following sunspots (\opencite{vanDriel90}), while the following sunspots tend to be affected by turbulent motions in the surroundings which cause the fragmentation of the flux tube (\opencite{Fan93}).
Therefore we mainly measured the circulation of the preceding sunspots and adopted their handedness of twist as more reliable representatives of the ARs.
To decide the sign of writhe, we considered the evolution of the $\delta$-part as described in Section~\ref{Tilt angle}.

The flare that corresponded to the peak X-ray flux represents only a single event of an AR through the observational period and the event did not always occur in the $\delta$-state.
On the other hand, FI can be measured over the definite period of the $\delta$-duration.
Therefore, we regarded FI as a better parameter of flare activity in the $\delta$-state than XR$_{\rm max}$.

\setlength{\tabcolsep}{3pt}  
\begin{table}[H]
\begin{flushleft}
\caption{
Summary of parameters for 31ARs studied.
The upper part of the table is for ARs in the northern hemisphere, while the lower part is for those in the southern hemisphere.
Here, XR$_{\rm max}$ means the peak X-ray flux in the observed period of each AR.
Flare Index indicates the value integrated over the observed period (the value integrated over the $\delta$-state is shown in parentheses).
$S_{\rm max}$ means the maximum sunspot area in the observed period.
$S_{\rm Umax}$ means the maximum umbral area in the observed period (the value in the $\delta$-state is shown in parentheses).
$\delta$-term shows the integrated duration of the $\delta$-state.
The twist and writhe are indicated by R (right-handed) or L (left-handed) .
The ``Type'' column show the emergence type of ARs; TT (top-to-top), DK (downward knotted), UK (upward knotted), and QB (quasi-$\beta$), respectively.
}
\begin{tabular}{c r r r r r c c c}
\hline
\raisebox{1.8ex}{AR} & \shortstack{XR$_{\rm max}$\\(Wm$^{-2}$)} & \raisebox{1.8ex}{Flare Index} & \shortstack{$S_{\rm max}$\\(Mm$^2$)} & \shortstack{$S_{\rm Umax}$\\(Mm$^2$)} & \shortstack{$\delta$-\\term\\(h)} & \raisebox{1.8ex}{Twist} & \raisebox{1.8ex}{Writhe} & \raisebox{1.8ex}{Type}\\
\hline
9165 & 59.0 & 257.0~~~(42.2) & 1851 & 291 (280) & ~24 & L & L & UK\\
9511 & 120.0 & 275.9~~(247.5) & 714 & 114 (114) & ~20 & R & R & DK\\
9678 & 20.0 & 103.1~~~(40.3) & 3046 & 623 (584) & ~26 & L\,to\,R & L & DK\\
9901 & 40.0 & 53.1~~~(53.1) & 2003 & 283 (283) & 188 & --- & R & ---\\
10412 & 9.8 & 22.8~~~~~(6.0) & 944 & 170 (160) &~27 & R & R & DK\\
10488 & 19.0 & 191.8~~(191.8) & 6786 & 1789 (1789) & 158 & L & L & DK\\
10564 & 110.0 & 219.0~~(209.0) & 3716 & 659 (659) & 110 & --- & --- & TT\\
10696 & 250.0 & 1114.9\,(1062.1) & 3841 & 829 (829) & 192 & L & L & TT\\
10720 & 710.0 & 2299.3\,(2299.3) & 6906 & 2147(2147) & 213 & L & L & TT\\
10956 & 2.9 & 3.9~~~~~(2.9) & 1054 & 151 (151) &112& --- & --- & ---\\
\hline
8506 & 3.3 & 9.1~~~~(0.0) & 1285 & 245 (211) & ~8 & R & --- & QB\\
8926 & 2.3 & 96.6~~~(41.6) & 805 & 152 (152) & ~52 & L & L & DK\\
9494 & 10.0 & 59.9~~~(26.7) & 1105 & 251 (159) & ~32 & R & R & DK\\
9775 & 22.0 & 115.3~~~(11.0) & 1415 & 144 (144) & ~32 & L & L & DK\\
9900 & 4.4 & 11.2~~~~~(0.0) & 633 & 75 ~ (67) & ~10 & --- & --- & ---\\
9904 & 9.2 & 8.2~~~~~(0.0) & 247 & 52 ~~(52) & ~30 & --- & L & ---\\
9906 & 26.0 & 58.1~~~\,(48.5) & 2907 & 540 (540) & 114 & R & R & DK\\
10017 & 150.0 & 282.5~~(174.4) & 2298 & 341 (341) & 34 & R & R & TT\\
10050 & 8.6 & 57.5~~~~~(6.5) & 3408 & 627 (600) & 38 & --- & --- & QB\\
10119 & 8.8 & 54.9~~~(14.0) & 3192 & 541 (541) & 92 & L & --- & QB\\
10137 & 40.0 & 174.7~~(170.0) & 1330 & 252 (252) & 88 & L & L & DK\\
10226 & 68.0 & 231.6~~(124.0) & 2812 & 578 (578) & 110 & L & R & DK\\
10314 & 150.0 & 511.1~~(443.9) & 2367 & 635 (635) & 94 & R\,to\,L & R & DK\\
10417 & 5.2 & 39.7~~~~~(6.9) & 1405 & 199 (158) & ~15 & --- & --- & QB\\
10456 & 4.4 & 6.6~~~~~(2.2) & 772 & 101 ~(60) & ~14 & --- & --- & QB\\
10551 & 3.2 & 6.9~~~~~(0.0) & 1531 & 221 (221) & ~38 & --- & --- & QB\\
10591 & 12.0 & 12.0~~~\,(12.0) & 412 & 82 ~~(82) & ~32 & R & --- & ---\\
10798 & 56.0 & 84.8~~~\,(84.3) & 2931 & 461 (461) & ~65 & R & --- & TT\\
10826 & 78.0 & 204.7~~(200.9) & 2152 & 390 (390) & ~56 & L & L & DK\\
10848 & 41.0 & 14.9~~~~~(0.0) & 1466 & 181 (181) & ~21 & --- & --- & ---\\
10865 & 14.0 & 30.0~~~~~(0.0) & 2532 & 568 (285) & ~38 & --- & --- & QB\\
\hline
\end{tabular}
\end{flushleft}
\label{ARs_property}
\end{table}

\begin{figure}[H] 
\begin{center}
\includegraphics[scale=0.36,angle=0,clip,trim= 0 0 0 0]{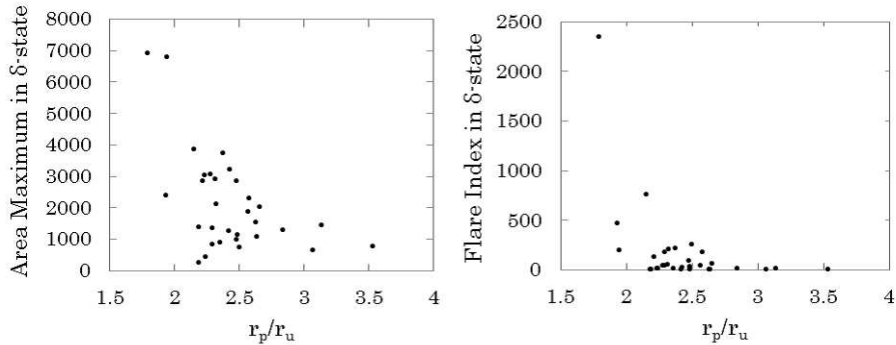} 
\caption{
$S_{\rm max}$ (left) and FI (right) {\it versus} the penumbral-umbral radius ratio in the $\delta$-state.
}
\label{PU_ratio}
\end{center}
\end{figure}

\subsubsection{Flare Activity and Sunspot Area}
\label{Flare and Area}


\inlinecite{Sammis00} statistically studied the relation between the flare activity and the maximum sunspot area for the ARs in Cycle 22, and found that ARs with larger maximum sunspot area produced higher peak X-ray flux.
Especially they showed that $\beta\gamma\delta$-ARs had the largest maximum sunspot area and hence produced the strongest flares.
Now we will check the relation between the FI (the proxy of flare activity level) and the maximum umbral area for our sampled $\beta\gamma\delta$-ARs in Cycle 23.  

\begin{figure}[t] 
\begin{center}
\includegraphics[scale=0.49,angle=0,clip,trim= 10 0 5 0]{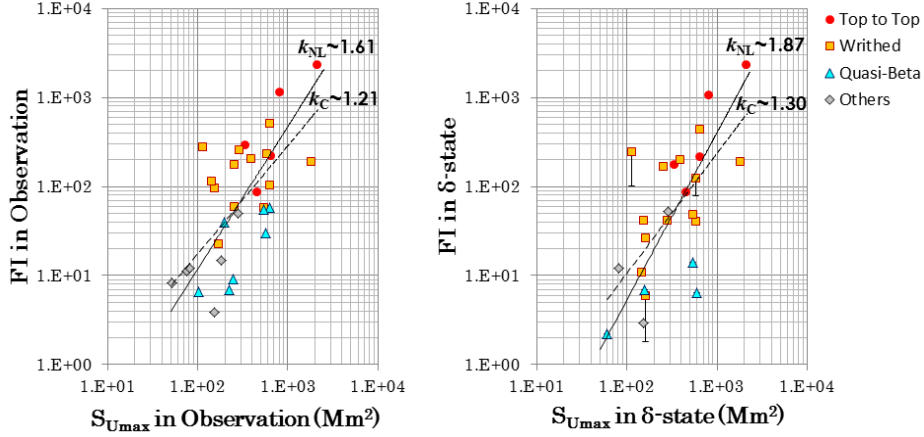}  
\caption{
Flare index {\it versus} maximum umbral area in a logarithmic scale.
The left panel is for the data in the observed period and the right panel is for those in the $\delta$-state.
Six ARs which showed no flare activity in the $\delta$-state were excluded.
Different symbols indicate ``top-to-top" (red circles), ``writhed" (orange squares, both DK and UK types in Table 1), ``quasi-$\beta$" (blue triangles), and ``others" (gray diamonds), respectively.
The error-bars were assigned by considering the effect of data gaps (see text).
The results of non-linear least-squares fitting (power-law index $k_{\rm NL}$) and classical least-squares fitting (power-law index $k_{\rm C}$) are shown by the solid and the dashed lines, respectively.
}
\label{FI_Umbra}
\end{center}
\end{figure}

First, we confirm that the relation found by \inlinecite{Sammis00} also holds in our data.
The physical meaning of the relation is that the sunspot area is a measure of the total magnetic flux of the region and hence represents the magnetic energy content of the region.
The ARs with large content of magnetic energy can produce large release of energy as flaring.
As we do not have the magnetic flux values for all the ARs in Cycle 23, we have to use the measured area as a proxy parameter for the magnetic flux.
The line-of-sight magnetic data by SOHO, for example, cannot provide a reliable estimate of the magnetic flux, without the correction for the projection effect and the field inclination to the vertical.
We took the umbral area as a better proxy parameter of the magnetic flux than the total sunspot area including the penumbral area, because of two reasons.
The first is that larger sunspots have smaller contribution of penumbra to the total sunspot area (left panel of Figure~\ref{PU_ratio}) and contain more magnetic flux in the umbral area (\opencite{Antalova91}; \opencite{Jin06}).
The other empirical reason is that FI does not depend so strongly on the penumbra-umbra radius ratio when this ratio is high (right panel of Figure~\ref{PU_ratio}). 

Next, we notice large dispersion in FI in Figure~\ref{FI_Umbra}.
For a given maximum umbral area, there appears a variety of ARs from highly flare productive ones to less productive ones.
In Figure~\ref{FI_Umbra}, $\beta\gamma\delta$ ARs are roughly split into different emergence types, suggesting that the emergence mode is another key parameter to characterize the flare productivity. 
By comparing the two panels in the figure, we can see that the separation of the three groups in the right panel is clearer than in the left panel. 
Since the clear separation into groups may give hints of the $\delta$-state, we will mainly concentrate to the flare activity and the area variation in the $\delta$-state hereafter.

\subsubsection{Flare Activity and Emergence Type}
\begin{figure}[H] 
\begin{center}
\includegraphics[scale=0.5,angle=0,clip,trim= 13 0 40 5]{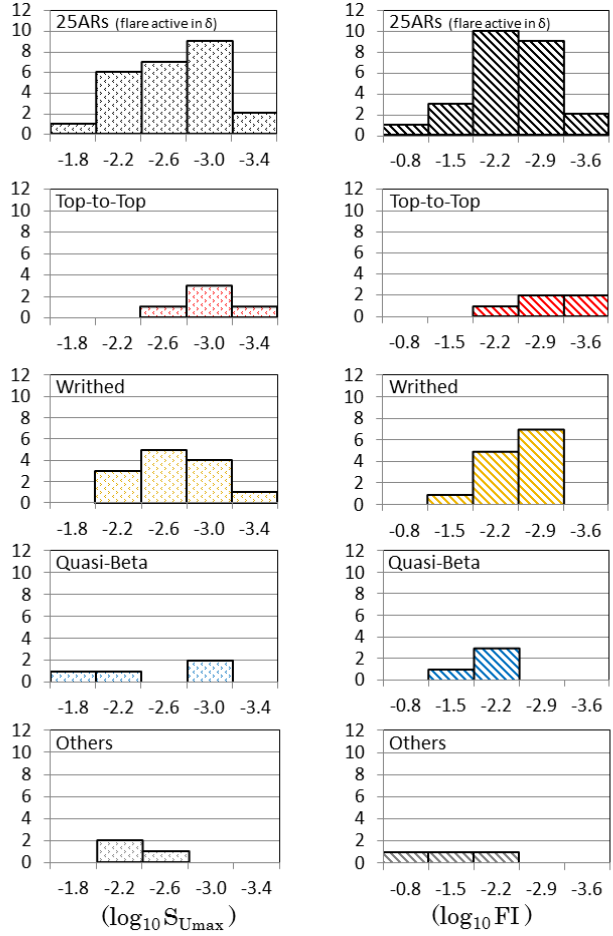}
\caption{
Histograms of umbral area maximum (left) and FI (right) in the $\delta$-state period.
From the second row downward are shown ``top-to-top", ``writhed", ``quasi-$\beta$", and ``others" types, respectively, and the top row shows the total.
}
\label{Area_FI_histo}
\end{center}
\end{figure}

Figure~\ref{Area_FI_histo} shows the histogram of  $S_{\rm Umax}$ and FI distributions grouped according to the emergence type in the $\delta$-state.
We combined the groups DK (downward-kinked) and UK (upward-kinked) in Table 1 as the ``writhed" type. 
The bin sizes of $S_{\rm Umax}$ and FI are $10^{0.4} (\approx 2.51)$ and $10^{0.7} (\approx 5.01)$, respectively.
The grouping follows the decreasing order of magnetic complexity ({\it i.e.} ``top-to-top'', ``writhed'', ``quasi-$\beta$" and ``others'').
As the topological complexity decreases, both $S_{Umax}$ and FI distribute to lower values.
The relation found here clearly indicates that the flare activity of $\beta\gamma\delta$ ARs depends not only on the umbral area but also on the magnetic complexity.
Moreover, it suggests that the large total content of magnetic flux represented by the umbral area is a necessary condition and significant magnetic complexity may be another more stringent necessary condition for the activation of strong flares.     

\subsubsection{Twist and Writhe of $\delta$-Spots}
When we classified the 31 ARs by their emergence mode, we found 5 ARs as ``top-to-top", 13 as ``writhed", 7 as ``quasi-$\beta$" and 6 as ``others". 
Although the ``quasi-$\beta$" and ``others" groups hardly ever showed the signs of twist and writhe, the ``top-to-top" and ``writhed" ARs clearly showed the twist and writhe.
In most cases of the ``writhed" type, the signs of twist and writhe agree with each other at least in their initial phase (12 cases out of 13).
This result strongly suggests that the formation of the $\delta$-spot is not due to the gathering of disjointed EFRs but due to the emergence of a singly connected structure (left panel in Figure~\ref{topological model}).
The``top-to-top" ARs also show high probability of agreement in their helicity signs (three cases out of five).
We speculate that the ``top-to-top" type has a more developed writhing structure than the ``writhed" type (right panel of Figure~\ref{topological model}).
However,  according  to the study of \inlinecite{Park12}, some ``top-to-top" ARs may be associated with the injection of opposite-sign helicities through flux emergence. 
Therefore the model of ``top-to-top" ARs  is at present controversial and further investigation is necessary to finalize the model.

The majority of ``writhed" ARs were of DK type (12 cases out of 13).
The UK type was quite rare.
The dominance of DK seems natural because a simple writhed magnetic tube with upward kink will emerge through the photosphere as a normal $\beta$-type magnetic distribution and will rarely be identified as the $\delta$-type sunspots.

\subsubsection{$\delta$-Duration and Emergence Type}
\label{delta-duration}
As shown in the left panel of Figure~\ref{delta_FA}, more magnetically complex ARs have longer $\delta$-duration.
We already confirmed that more magnetically complex ARs have larger maximum umbral area in Figure~\ref{FI_Umbra}.
   
In general, larger sunspots have longer lifetime (\opencite{Gnevyshev38}; \opencite{Waldmeier55}).
Recurrent sunspot groups are found to exhibit a slightly larger lifetime than that based on the Gnevyshev-Waldmeier relationship by \inlinecite{Petrovay97}.
It is also well known that larger sunspots tend to be darker and have stronger magnetic fields (\opencite{Solanki02}; \opencite{Livingston02}).
Furthermore, \inlinecite{Shi94} reported that the productivity of $\delta$-spots for X-class flares is closely correlated to the lifetime of $\delta$-spots.
Our result suggests that the long lifetime of active $\delta$-spots is closely related to their magnetic complexity (twist and writhe).
Although magnetic tubes are eventually dispersed due to the action of turbulent convection (\opencite{Petrovay_Moreno97}) and/or fragmentation by fluting instabilities, the twisted tubes are more resistant against these dissolving actions.
It seems difficult that the magnetic lines of force of the tube above the solar surface will be torn off by convective motions in lower layers, as the lines of force are entangled and hooked to the magnetic tube in the corona.
Further, the twist will suppress the fluting instability of magnetic tubes (\opencite{Priest14}).
Threfore, we argue that the longevity of an active $\delta$-spot depends on its magnetic twist.

\subsubsection{Possible Power-law Relations among Parameters}
\label{power-law}
On every scatter plot (Figures~\ref{FI_Umbra} and \ref{delta_FA}), we performed the power law fitting to the data. Two methods were used to derive the power-law index $k$.
The first is the classical least-squares fitting (C-LS) to log-log plots and we obtain the power-law index $k_{\rm C}$.
The second is the non-linear least-squares fitting (NL-LS) method based on the Gauss-Newton algorithm ({\it cf.} \opencite{Hansen12}), leading to $k_{\rm NL}$. For the data points in the left panel of Figure~\ref{FI_Umbra} ($S_{\rm Umax}$ {\it vs.} FI in the observed period), we found $k_{\rm C} =1.21 \pm 0.24$ and $k_{\rm NL}=1.61 \pm 0.33$.
In the right panel of Figure~\ref{FI_Umbra} ($S_{\rm Umax}$ {\it vs.} FI in the $\delta$-state), we found $k_{\rm C}=1.30 \pm 0.31$ and $k_{\rm NL}=1.87 \pm 0.50$. FI follows a power-law dependence on $S_{\rm Umax}$ with an index around 2, especially in the $\delta$-state.
It is worthwhile to note that \inlinecite{Magara14} discussed a power-law relationship between the magnetic free energy and the total magnetic flux for twisted emerging loops with an index of 2 from his numerical simulation.

We next show the relation between the $\delta$-duration and FI in the $\delta$-state (left panel of Figure~\ref{delta_FA}).
Although we could not get the converged solution of NL-LS, we obtained $k_{\rm C}= 1.54 \pm 0.34$ when we excluded the ``others'' group as they have large dispersion. For the relation between the $\delta$-duration and $S_{\rm Umax}$ in the $\delta$-state (right panel of Figure~\ref{delta_FA}), we found $k_{\rm C}=0.76 \pm 0.16$ and $k_{\rm NL}=1.07 \pm 0.28$. 
Combining the last two relations, namely both plots of Figure~\ref{delta_FA}, we can infer a power-law relation FI $\sim S_{\rm Umax}^{\ \,2.0}$ from C-LS which is consistent with the result derived in the previous paragraph.

\begin{figure}[t] 
\begin{center}
\includegraphics[scale=0.38,angle=0,clip,trim= 10 0 5 0]{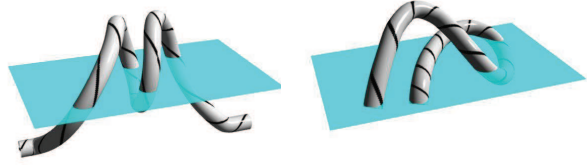} 
\caption{
Schematic models for $\beta\gamma\delta$ ARs. The left panel shows a general model for the downward-knotted type such as NOAA 10826.
The right panel shows a possible model for the ``top-to-top" type such as NOAA 10720.
}
\label{topological model}
\end{center}
\end{figure}

\begin{figure}[H] 
\begin{center}
\includegraphics[scale=0.47,angle=0,clip,trim= 0 0 0 0]{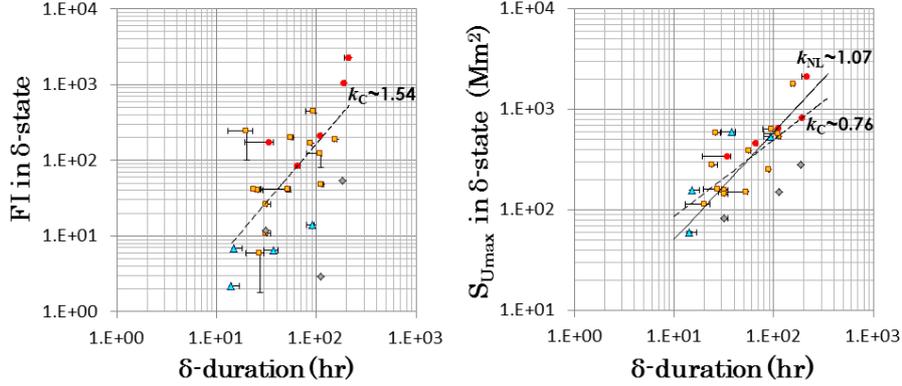}  
\caption{
Flare Index (left) and maximum umbral area (right) in the $\delta$-state as a function of the $\delta$-duration.
The symbols are the same as in Figure~\ref{FI_Umbra}. Six ARs which showed no flare activity in the $\delta$-state were excluded.
The error-bars were assigned by considering the effect of data gaps (see text).
The results of non-linear least-squares fitting (power-law index $k_{\rm NL}$) and classical least-squares fitting (power-law index $k_{\rm C}$) are shown by the solid and the dashed lines, respectively.
}.
\label{delta_FA}
\end{center}
\end{figure}

\section{Discussion} 
      \label{S-Discussion}   

By paying attention to the initial phase of $\beta\gamma\delta$ ARs in Cycle 23, we investigated what are the important configurations which lead to the flare activity.
From our investigation, we discovered three representative topological types which have strong correlations with the flare activity, namely ``quasi-$\beta$'',``writhed'', and ``top-to-top''.
We confirmed that the ARs with higher topological complexity show stronger flare activity.

It is interesting to note that almost all ARs classified as the ``writhed" type are of DK type.
Once the downward knot structure is formed, a dense plasma ``reservoir" is formed in the knot and the whole structure of the flux tube may become stable owing to the effect of the mass as an anchor (left panel of Figure~\ref{topological model}). 
Under a different situation, \inlinecite{Magara03} proposed a similar idea in their numerical simulation of the emergence of a twisted flux tube in the convection zone. 
Subsequently, \inlinecite{Magara11} has studied the case of the emergence of a U-loop formed below the photosphere.
\inlinecite{vanDriel00} have reported that the emergence of a U-loop which connected the opposite polarity legs of two $\Omega$-loops below the photosphere. 
Recently, \inlinecite{Poisson13} studied the possibility of a downward kinked structure in the mid part of NOAA10314.  
Thus the idea of DK structure is not so exceptional and applicable to twisted magnetic tubes, with several evidence.
The DK structure as an anchor may contribute to the stability of the AR.

Subsequently we checked the case of  the ``top-to-top'' group.
They show clear twist and writhe structures with common handedness (three cases out of five). 
Therefore, we infer that the three-dimensional collision is not an accidental one but is attributed to the extreme development of a  writhing structure.
For example NOAA10696, classified as the ``top-to-top'' type, had the second strongest FI in our sampled ARs and showed both characters of three-dimensional collision and downward knotted structure.
The AR is namely a hybrid type of ``writhed'' and ``top-to-top''.
Such a configuration of this AR reminds us that there may be continuity of developments between the ``top-to-top'' and ``writhed'' structures.
We propose a schematic model for NOAA10720 with a highly writhed structure as shown in Figure~\ref{topological model} (right panel). 
We can point out that a similar, highly writhed model for NOAA 7926 was proposed by \inlinecite{Pevtsov98}.     

\inlinecite{Brown03} studied the rotation of seven sunspots using high resolution white light images with TRACE.
They proposed that a possible mechanism for sunspot rotation is the emergence of a pre-twisted flux tube.  
On the other hand, \inlinecite{Yan09} claimed that the rotation of sunspots originates from the photospheric flows, based on their study of combined photospheric and coronal observations.
In our study, the rotation of sunspots themselves and the tilt angle rotation of EFR are rarely observed in small and flare-inactive ARs.  
If the rotation is mainly attributed to photospheric flows, rotational motions of sunspots will be observed commonly regardless of the sunspot size. 
However, as shown in our statistical study, the ARs of prominent twist and writhe had larger area and showed stronger flare activity among the sampled ARs.
In addition, as described in Section~\ref{delta-duration}, the pre-twisted flux tube model is in harmony with several features of $\delta$-spots.
Therefore we conclude that the origin of sunspot rotation is the emergence of the pre-twisted flux tube.

It is known that twist and writhe helicities transform each other in an ideal thin flux tube (\opencite{Calugareanu59}; \opencite{Moffat92}; \opencite{Ricca95}; \opencite{Torok10}).
In this study, we adopted a picture in that the initial twist of ARs will be mostly generated by turbulent flows in the convection zone. 
When the twist increases in an AR, the twist transforms into the writhe in the middle part of the flux tube by the kink instability (\opencite{Torok10}).
Then it is natural to find the sign relationship as seen in this study.
The previous studies have shown that the kink instability facilitates the formation of  $\delta$-spots ({\it e.g.} \opencite{Linton99}).
\inlinecite{Tian05} pointed out that the signs of twist and writhe tend to agree in well-developed ARs under kink instability.

Regarding the classification of $\delta$-spots, we review the relations between the types classified by our study and the types by \inlinecite{Zirin87} as introduced in Section 1.
We can consider that the first type of Zirin and Liggett (ZL) is the same as the ``top-to-top" type of ours.
The third type of ZL may be equal to the ``writhed" group, although we believe that the configuration is not formed by different dipoles but by a connected structure below the photosphere.
However, the second type of ZL was seldom observed in our study.
This type was seen in only one or two cases in the ``others" group of our study.
This type may be more likely to occur in recurrent sunspots rather than in sunspots in their young stage studied here.
The ``quasi-$\beta$" type was not considered in the ZL classification as they may not produce notable flares.   

We discussed the possible power-law relations among the $\beta\gamma\delta$-spot parameters.
Our physical interpretation is as follows.
FI is a good proxy of the total energy released in flares in an AR and thus is a measure of the total magnetic free energy of the region supplied by the gas flow field.
The maximum umbral area $S_{\rm Umax}$ is a measure of the magnetic flux of the region.
The functional relation of a power law sometimes is interpreted as the evidence of a scale-free process (\opencite{Shimizu95}; \opencite{Shimojo99}; \opencite{Nishizuka09}).
In our case, the total free energy supply or injection can be considered as  due to a kind of scale-free processes, probably the helicity injection process by the turbulent convective flows or the differential rotation.
The process seems to work independent of the total amount of magnetic flux.
The difference in FI among the emergence-type groups may be the difference in topology of the magnetic tubes.
While the ``quasi-$\beta$''\,type may have a simple form with less twist and less free-energy, the ``writhed''\,type may have a singly knotted or kinked tube with more free-energy, and the ``top-to-top''\,type may have a doubly or multiply knotted tube with much more free-energy. 

\section{Conclusions} 
      \label{S-Conclusions}   

We have studied the initial evolution of 31 $\beta\gamma\delta$ ARs and derived the following conclusions.\\
\romannumeral1 ) From the point of view of topology, emerging $\beta\gamma\delta$ ARs can be classified broadly into three categories: ``quasi-$\beta$", ``writhed", and ``top-to-top".
The ``top-to-top" type has the most complex topology and the ``quasi-$\beta$" type is the simplest type.
The ``Writhed" group has medium complexity among the three types.\\
\romannumeral2 ) ARs of more complex topology tend to exhibit higher flare activity.\\ 
\romannumeral3 ) The signs of twist and writhe both tend to be consistent with each other in the ``writhed" type.\\
\romannumeral4 ) The downward-knotted structure in the mid portion of the flux tube is the essential element of active $\beta\gamma\delta$ ARs.\\
\romannumeral5 ) The flare activity of $\beta\gamma\delta$ ARs are highly correlated not only with the sunspot areas but also with the magnetic complexity.\\ 
\romannumeral6 ) There is a possible scaling-law between FI and $S_{\rm Umax}$.\\ 

Finally, we will give some comments on the future extension of our study.
Direct measurements of magnetic helicity will be essential to study the free-energy accumulation in ARs.
Observations of the ARs throughout their lifetime along with stereoscopic views from space will be desirable.
Further it will be an important challenge to diagnose the magnetic topology of ARs in the convection zone by improving local helioseismology.


\clearpage
\begin{acks}
The authors appreciate useful comments from an anonymous referee.
We thank the SOHO, TRACE, and GOES consortia for their data.
SOHO is a international cooperation between ESA and NASA.
We are grateful to the staff of Kwasan and Hida observatories of Kyoto University, for their continuing help. 
We wish to thank Hiroki Kurokawa and Yin Zhang for useful discussion.
This work was supported by the Grant-in-Aid for the Global COE Program ``The Next Generation of Physics, Spun from Universality and Emergence'' from the Ministry of Education, Culture, Sports, Science and Technology (MEXT) of Japan.
We are partially supported by the grant-in-aid from the Japanese MEXT (PI: R.Kitai No.26400235).  
\end{acks}



\begin{thebibliography}{}
\bibitem[\protect\citeauthoryear{Abramenko}{2005}]{Abramenko05}
Abramenko, V. I. : 2005, \apj{}, \textbf{629}, 1141.
\adsurl{2005ApJ...629.1141A}
\doiurl{10.1086/431732}
\bibitem[\protect\citeauthoryear{Antalov\'a}{1991}]{Antalova91}
Antalov\'a, A.: 1991, \textit{Bull. Astron. Inst. Czechosl.} \textbf{42}, 316.
\adsurl{1991BAICz..42..316A}
\bibitem[\protect\citeauthoryear{Antalov\'a}{1996}]{Antalova96}
Antalov\'a, A.: 1996, \textit{Contrib. Astron. Obs. Skalnate Pleso} \textbf{26}, 98.
\adsurl{1996CoSka..26...98A}
\bibitem[\protect\citeauthoryear{Brown \etal}{2003}]{Brown03}
Brown, D. S., Nightingale, R. W., Alexander, D., Schrijver, C. J., Metcalf, T. R., Shine, R. A., Title, A. M., Wolfson, C. J.: 2003, \solphys{} \textbf{216}, 79.
\adsurl{2003SoPh..216...79B}
\doiurl{10.1023/A:1026138413791}
\bibitem[\protect\citeauthoryear{C\u{a}lug\u{a}reanu}{1959}]{Calugareanu59}
C\u{a}lug\u{a}reanu, G.: 1959, {\it Rev. Roum. Math. Pure Appl.} \textbf{4}, 5.
\bibitem[\protect\citeauthoryear{Chae}{2001}]{Chae01}
Chae, J.: 2001, \apj{} \textbf{560}, 95.
\adsurl{2001ApJ...560L..95C}
\doiurl{10.1086/324173}
\bibitem[\protect\citeauthoryear{Chae \etal}{2001}]{Chae_etal01}
Chae, J., Wang, H., Qiu, J., Goode, P.R., Strous, L., Yun, H. S.: 2001,\apj{} \textbf{560}, 476.
\adsurl{2001ApJ...560..476C}
\doiurl{10.1086/322491}
\bibitem[\protect\citeauthoryear{Delaboudini\'ere \etal}{1995}]{Delabou95}
Delaboudini\'ere, J. -P., Artzner, G. E., Brunaud, J., Gabriel, A. H., Hochedez, J. F., Millier, F.,
\textit{et al.}: 1995, \solphys{} \textbf{162}, 291.
\adsurl{1995SoPh..162..291D}
\doiurl{10.1007/BF00733432}
\bibitem[\protect\citeauthoryear{Domingo \etal}{1995}]{Domingo95}
Domingo, V., Fleck, B., Poland, A. I.: 1995, \solphys{} \textbf{162}, 1.
\adsurl{1995SoPh..162....1D}
\doiurl{10.1007/BF00733425}
\bibitem[\protect\citeauthoryear{Fan \etal}{1993}]{Fan93}
Fan, Y., Fisher, G. H., Deluca, E. E.: 1993, \apj{} \textbf{405}, 390.
\adsurl{1993ApJ...405..390F}
\doiurl{10.1086/172370}
\bibitem[\protect\citeauthoryear{Fan \etal}{1999}]{Fan99}
Fan, Y., Zweibel, E. G., Linton, M. G., Fisher, G. H.: 1999, \apj{} \textbf{521}, 460.
\adsurl{1999ApJ...521..460F}
\doiurl{10.1086/307533}
\bibitem[\protect\citeauthoryear{Gnevyshev}{1938}]{Gnevyshev38}
Gnevyshev, M.N.: 1938, {\it Pulkovo Obs. Circ.} \textbf{24}, 37.
\bibitem[\protect\citeauthoryear{Guo \etal}{2013}]{Guo13}
Guo, Y., Ding, M. D., Cheng, X., Zhao, J. Pariat, E.: 2013, \apj{} \textbf{779}, 157.
\adsurl{2013ApJ...779..157G}
\doiurl{10.1088/0004-637X/779/2/157}
\bibitem[\protect\citeauthoryear{Handy \etal}{1999}]{Handy99}
Handy, B. N., Acton, L. W., Kankelborg, C. C., Wolfson, C. J., Akin, D. J., Bruner, M. E.,
 \textit{et al.}: 1999, \solphys{} \textbf{187}, 229.
\adsurl{1999SoPh..187..229H}
\doiurl{10.1023/A:1005166902804}
\bibitem[\protect\citeauthoryear{Hansen \etal}{2012}]{Hansen12}
Hansen, P. C., Pereyra, V., Scherer, G.,: 2012, {\it Least Squares Data Fitting With Applications}, Johns Hopkins Univ. Press, Baltimore, MD, 163-175
\bibitem[\protect\citeauthoryear{Ikhsanov and Marushin}{2003}]{Ikhsanov03}
Ikhsanov, R. N., Marushin, Y. V.: 2003, \textit{arXiv:astro-ph} 0311114.
\adsurl{2003astro.ph.11114I}
\bibitem[\protect\citeauthoryear{Ikhsanov \etal}{2004}]{Ikhsanov04}
Ikhsanov, R. N., Marushin, Yu. V., Ikhsanov, N. R.: 2004, In: Stepanov, A.V., Benevolenskaya, E. E., Kosovichev, A. G. (eds.), {\it Multi-Wavelength Investigations of Solar Activity, IAU Symp.} \textbf{223}, 257.
\adsurl{2004IAUS..223..257I}
\doiurl{10.1017/S1743921304005812}
\bibitem[\protect\citeauthoryear{Ishii \etal}{1998}]{Ishii98}
Ishii, T. T., Kurokawa, H., Takeuchi, T. T.: 1998, \apj{} \textbf{499}, 898.
\adsurl{1998ApJ...499..898I}
\bibitem[\protect\citeauthoryear{Jin \etal}{2006}]{Jin06}
Jin, C. L.; Qu, Z. Q., Xu, C. L., Zhang, X. Y., Sun, M. G.: 2006, {\it Astrophys.  Space Sci.} \textbf{306}, 23.
\adsurl{2006Ap&SS.306...23J}
\doiurl{10.1007/s10509-006-9217-6}
\bibitem[\protect\citeauthoryear{Jing \etal}{2006}]{Jing06}
Jing, J., Song, H., Abramenko, V. I., Tan, C., Wang, H.: 2006, \apj{} \textbf{644}, 1273.
\doiurl{10.1086/503895}
\adsurl{2006ApJ...644.1273J}
\bibitem[\protect\citeauthoryear{Joshi and Joshi}{2004}]{Joshi04}
Joshi, B., Joshi, A.: 2004, \solphys{} \textbf{219}, 343.
\adsurl{2004SoPh..219..343J}
\doiurl{10.1023/B:SOLA.0000022977.95023.a7}
\bibitem[\protect\citeauthoryear{K\"unzel}{1960}]{Kunzel60}
K\"unzel, H.: 1960, {\it Astron. Nachr.} \textbf{285}, 271.
\adsurl{1960AN....285..271K}
\bibitem[\protect\citeauthoryear{Kurokawa}{1987}]{Kurokawa87}
Kurokawa, H.: 1987, \solphys{} \textbf{113}, 259.
\adsurl{1987SoPh..113..259K}
\doiurl{10.1007/BF00147706}
\bibitem[\protect\citeauthoryear{Kurokawa \etal}{2002}]{Kurokawa02}
Kurokawa, H. Wang, T., Ishii, T. T.: 2002, \apj{} \textbf{572}, 598.
\adsurl{2002ApJ...572..598K}
\doiurl{10.1086/340305}
\bibitem[\protect\citeauthoryear{Leka \etal}{1996}]{Leka96}
Leka, K. D., Canfield, R. C., McClymont, A. N., van Driel-Gesztelyi, L.: 1996, \apj{} \textbf{462}, 547. 
\adsurl{1996ApJ...462..547L}
\doiurl{10.1086/177171}
\bibitem[\protect\citeauthoryear{Linton \etal}{1999}]{Linton99}
Linton, M. G., Fisher, G. H., Dahlburg, R. B., Fan, Y.: 1999, \apj{} \textbf{522}, 1190.
\adsurl{1999ApJ...522.1190L}
\doiurl{10.1086/307678}
\bibitem[\protect\citeauthoryear{Livingston}{2002}]{Livingston02}
Livingston, W.: 2002, \solphys{}, {\textbf 207}, 41.
\adsurl{2002SoPh..207...41L}
\doiurl{10.1023/A:1015555000456}
\bibitem[\protect\citeauthoryear{L\'opez Fuentes \etal}{2000}]{Lopez00}
L\'opez Fuentes, M. C., Demoulin, P., Mandrini, C. H., van Driel-Gesztelyi, L.: 2000, \apj{} \textbf{544}, 540.
\adsurl{2000ApJ...544..540L}
\doiurl{10.1086/317180}
\bibitem[\protect\citeauthoryear{L\'opez Fuentes \etal}{2008}]{Lopez08}
L\'opez Fuentes, M. C., Mandrini, C. H.: 2008, \textit{Bol. Asoc. Argent. Astron.} \textbf{51}, 31.
\adsurl{2008BAAA...51...31L}
\bibitem[\protect\citeauthoryear{Luoni \etal}{2011}]{Luoni11}
Luoni, M.L., D{\'e}moulin, P., Mandrini, C. H., van Driel-Gesztelyi, L.: 2011, \solphys{} \textbf{270}, 45. 
\adsurl{2011SoPh..270...45L}
\doiurl{10.1007/s11207-011-9731-8}
\bibitem[\protect\citeauthoryear{Magara and Longcope}{2003}]{Magara03}
Magara, T, Longcope, D. W.: 2003, \apj{} \textbf{586}, 630. 
\adsurl{2003ApJ...586..630M}
\doiurl{10.1086/367611}
\bibitem[\protect\citeauthoryear{Magara}{2011}]{Magara11}
Magara, T.: 2011, \pasj{} \textbf{63}, 417. 
\adsurl{2011PASJ...63..417M}
\doiurl{10.1093/pasj/63.2.417}
\bibitem[\protect\citeauthoryear{Magara}{2014}]{Magara14}
Magara, T.: 2014, \pasjl{} \textbf{66}, L6.
\adsurl{2014PASJ...66L...6M}
\doiurl{10.1093/pasj/psu049}
\bibitem[\protect\citeauthoryear{Moffat and Ricca}{1992}]{Moffat92}
Moffatt, H. K., Ricca, R. L.: 1992, {\it Proc. Roy. Soc. London A} \textbf{439}, 411.
\doiurl{10.1098/rspa.1992.0159}
\bibitem[\protect\citeauthoryear{Nishizuka \etal}{2009}]{Nishizuka09}
Nishizuka, N., Asai, A., Takasaki, H., Kurokawa, H., Shibata, K.: 2009, \apjl{} \textbf{694}, L74.
\adsurl{2009ApJ...694L..74N}
\doiurl{10.1088/0004-637X/694/1/L74}
\bibitem[\protect\citeauthoryear{November and Simon}{1988}]{November88}
November, L. J., Simon, G. W.: 1988, \apj{} \textbf{333}, 427. 
\adsurl{1988ApJ...333..427N}
\doiurl{10.1086/166758}
\bibitem[\protect\citeauthoryear{Park \etal}{2012}]{Park12}
Park, J., Moon, Y.-J., Gopalswamy, N.: 2012, \apj{} \textbf{750}, 48.
\adsurl{2012ApJ...750...48P}
\doiurl{10.1088/0004-637X/750/1/48}
\bibitem[\protect\citeauthoryear{Petrovay and Moreno-Insertis}{1997}]{Petrovay_Moreno97}
Petrovay, K., Moreno-Insertis, F.: 1997, \apj{} \textbf{485}, 398.
\adsurl{1997ApJ...485..398P}
\bibitem[\protect\citeauthoryear{Petrovay and van Driel-Gesztelyi}{1997}]{Petrovay97}
Petrovay, K., van Driel-Gesztelyi, L.: 1997, \solphys{} \textbf{176}, 249.
\adsurl{1997SoPh..176..249P}
\doiurl{10.1023/A:1004988123265}
\bibitem[\protect\citeauthoryear{Pevtsov \etal}{1998}]{Pevtsov98}
Pevtsov, A. A., Longcope, D. W.: 1998, \apj{} \textbf{508}, 908.
\adsurl{1998ApJ...508..908P}
\doiurl{10.1086/306414}
\bibitem[\protect\citeauthoryear{Poisson \etal}{2013}]{Poisson13}
Poisson, M., L\'opez Fuentes, M., Mandrini, C. H., Demoulin, P., Pariat, E.: 2013, \textit{Adv. Space Res.} \textbf{51}, 1834.
\adsurl{2013AdSpR..51.1834P}
\doiurl{10.1016/j.asr.2012.03.010}
\bibitem[\protect\citeauthoryear{Priest}{2014}]{Priest14}
Priest, E.: 2014, {\it Magnetohydrodynamics of the Sun}, Cambridge University Press, Cambridge, 270.
\adsurl{2014masu.book.....P}
\bibitem[\protect\citeauthoryear{Ricca}{1995}]{Ricca95}
Ricca, R. L.: 1995, {\it J. Phys. A} \textbf{28}, 2335.
\adsurl{1995JPhA...28.2335R}
\doiurl{10.1088/0305-4470/28/8/024}
\bibitem[\protect\citeauthoryear{Rust and Kumar}{1996}]{Rust96}
Rust, D. M., Kumar, A.: 1996, \apjl{} \textbf{464}, 199.
\adsurl{1996ApJ...464L.199R}
\doiurl{10.1086/310118}
\bibitem[\protect\citeauthoryear{Sammis \etal}{2000}]{Sammis00}
Sammis, I., Tang, F., Zirin, H.: 2000, \apj{} \textbf{540}, 583.
\adsurl{2000ApJ...540..583S}
\doiurl{10.1086/309303}
\bibitem[\protect\citeauthoryear{Scherrer \etal}{1995}]{Scherrer95}
Scherrer, P. H., Bogart, R. S., Bush, R. I., Hoeksema, J. T., Kosovichev, A. G., Schou, J., {\it et al.}: 1995, \solphys{} \textbf{162}, 129.
\adsurl{1995SoPh..162..129S}
\doiurl{10.1007/BF00733429}
\bibitem[\protect\citeauthoryear{Shi and Wang}{1994}]{Shi94}
Shi, Z. X., Wang, J. X.: 1994, \solphys{} \textbf{149}, 105.
\adsurl{1994SoPh..149..105S}
\doiurl{10.1007/BF00645181}
\bibitem[\protect\citeauthoryear{Shimizu}{1995}]{Shimizu95}
Shimizu, T.: 1995, \pasj{} \textbf{47}, 251.
\adsurl{1995PASJ...47..251S}
\bibitem[\protect\citeauthoryear{Shimojo and Shibata}{1999}]{Shimojo99}
Shimojo, M., Shibata, K.:1999, \apj{} \textbf{516}, 934.
\adsurl{1999ApJ...516..934S}
\doiurl{10.1086/307156}
\bibitem[\protect\citeauthoryear{Solanki}{2002}]{Solanki02}
Solanki, S. K.: 2002, {\it Astron. Nachr.} \textbf{323}, 165.
\adsurl{2002AN....323..165S}
\doiurl{10.1002/1521-3994(200208)323:3/4<3C165::AID-ASNA165>3E3.0.CO;2-U}
\bibitem[\protect\citeauthoryear{Tanaka}{1975}]{Tanaka75}
Tanaka, K.:1975, \textit{BBSO Preprint No.0152}, Big Bear Solar Observatory
\bibitem[\protect\citeauthoryear{Tanaka}{1991}]{Tanaka91}
Tanaka, K.: 1991, \solphys{} \textbf{136}, 133.
\adsurl{1991SoPh..136..133T}
\doiurl{10.1007/BF00151700}
\bibitem[\protect\citeauthoryear{Tian \etal}{2005}]{Tian05}
Tian, L., Alexander, D., Liu, Y., Yang, J., 2005, \solphys{} \textbf{229}, 63.
\adsurl{2005SoPh..229...63T}
\doiurl{10.1007/s11207-005-3524-x}
\bibitem[\protect\citeauthoryear{T\"or\"ok \etal}{2010}]{Torok10}
T\"or\"ok, T., Berger, M. A., Kliem, B.: 2010, \aap{} \textbf{516}, 49.
\adsurl{2010A&A...516A..49T}
\doiurl{10.1051/0004-6361/200913578}
\bibitem[\protect\citeauthoryear{van Driel-Gesztelyi and Petrovay}{1990}]{vanDriel90}
van Driel-Gesztelyi, L., Petrovay, K.: 1990, \solphys{} \textbf{126}, 285.
\adsurl{1990SoPh..126..285V}
\doiurl{10.1007/BF00153051}
\bibitem[\protect\citeauthoryear{van Driel-Gesztelyi \etal}{1997}]{vanDriel97}
van Driel-Gesztelyi, L., Csepura, G., Schmieder, B., Malherbe, J.-M., Metcalf, T.: 1997, \solphys{} \textbf{172}, 151. 
\adsurl{1997SoPh..172..151V}
\doiurl{10.1023/A:1004975212949}
\bibitem[\protect\citeauthoryear{van Driel-Gesztelyi \etal }{2000}]{vanDriel00}
van Driel-Gesztelyi, L., Malherbe, J.-M., D\'emoulin, P.: 2000, \aap{} \textbf{364}, 845. 
\adsurl{2000A&A...364..845V}
\bibitem[\protect\citeauthoryear{Waldmeier}{1955}]{Waldmeier55}	
Waldmeier, M.: 1955, {\it Ergebnisse und Probleme der Sonnenforschung}, Geest \& Portig, Leipzig, 164-165
\adsurl{1955epds.book.....W},
\bibitem[\protect\citeauthoryear{Yan \etal }{2009}]{Yan09}
Yan, Xiao-Li., Qu, Zhong-Quan., Xu, Cheng-Lin., Xue, Zhi-Ke., Kong, De-Fang.: 2009, {\it Res. Astron. Astrophys.} \textbf{9}, 596.
\adsurl{2009RAA.....9..596Y}
\doiurl{10.1088/1674-4527/9/5/010}
\bibitem[\protect\citeauthoryear{Zirin and Liggett}{1987}]{Zirin87}
Zirin, H., Liggett, M. A.: 1987, \solphys{} \textbf{113}, 267.
\adsurl{1987SoPh..113..267Z}
\doiurl{10.1007/BF00147707}
\bibitem[\protect\citeauthoryear{Zwaan}{1978}]{Zwaan78}
Zwaan, C.: 1978, \solphys{} \textbf{60}, 213.
\adsurl{1978SoPh...60..213Z}
\doiurl{10.1007/BF00156523}
\end{thebibliography}

\end{article} 

\end{document}